\documentclass[sigconf]{acmart}

\acmConference[Conference 2024]{Conference on Software Engineering}{May 2024}{}

\usepackage{xspace}
\usepackage{multirow}
\usepackage{multicol}
\usepackage{booktabs}
\usepackage{threeparttable}
\usepackage{listings}
\usepackage{xcolor}
\usepackage{arydshln}
\usepackage{subcaption}
\usepackage{graphicx}
\usepackage{hyperref}
\usepackage{bbding}
\usepackage[skins]{tcolorbox}

\usepackage{times}
\usepackage{latexsym}

\usepackage[T1]{fontenc}
\usepackage[utf8]{inputenc}

\usepackage{microtype}

\definecolor{dkgreen}{rgb}{0,0.6,0}
\definecolor{gray}{rgb}{0.5,0.5,0.5}
\definecolor{mauve}{rgb}{0.58,0,0.82}
\definecolor{my-blue}{rgb}{0.98, 0.98, 1.0} %FAFAFF
\newcommand{\ourmethod}{\textit{DomCoder}\xspace}

\newcommand{\rev}[1]{\textcolor{black}{#1}}

\begin{document}

\title{On the Effectiveness of Large Language Models in Domain-Specific Code Generation}

%\author{Meng Chen \\
%    Shanghai Jiao Tong University
%    \texttt{mengchen@sjtu.edu.cn} \\\And
%    Hongyu Zhang \\
%    Chongqing University
%    \texttt{hyzhang@cqu.edu.cn} \\\And
%    Chengcheng Wan \\
%    East China Normal University
%    \texttt{ccwan@sei.ecnu.edu.cn} \\\And
%    Zhao Wei \\
%    Tencent Inc.
%    \texttt{zachwei@tencent.com} \\\And
%    Yong Xu \\
%    Tencent Inc.
%    \texttt{rogerxu@tencent.com} \\\And
%    Juhong Wang \\
%    Tencent Inc.
%    \texttt{julietwang@tencent.com} \\\And
%    Xiaodong Gu \\
%    Shanghai Jiao Tong University
%    \texttt{xiaodong.gu@sjtu.edu.cn} \\
%}

\author{Xiaodong Gu}
%\authornote{Corresponding author.}
\affiliation{%
  \institution{Shanghai Jiao Tong University}
  \city{Shanghai}
  \country{China}}
\email{xiaodong.gu@sjtu.edu.cn}

\author{Meng Chen}
%\authornotemark[1]
\affiliation{%
  \institution{Shanghai Jiao Tong University}
  \city{Shanghai}
  \country{China}
}
\email{mengchen@sjtu.edu.cn}

\author{Yalan Lin}
%\authornote{Both authors contributed equally to this research}
\affiliation{%
  \institution{Shanghai Jiao Tong University}
  \city{Shanghai}
  \country{China}
}
\email{linyalan@sjtu.edu.cn}

\author{Yuhan Hu}
%\authornotemark[1]
\affiliation{%
  \institution{Shanghai Jiao Tong University}
  \city{Shanghai}
  \country{China}
}
\email{suzhengv@sjtu.edu.cn}

\author{Hongyu Zhang}
\affiliation{%
  \institution{Chongqing University}
  \city{Chongqing}
  \country{China}}
\email{hyzhang@cqu.edu.cn}

\author{Chengcheng Wan}
\affiliation{%
  \institution{East China Normal University}
  \city{Shanghai}
  \country{China}
}
\email{ccwan@sei.ecnu.edu.cn}

\author{Zhao Wei}
\affiliation{%
 \institution{Tencent Inc.}
 \city{Beijing}
 \country{China}}
 \email{zachwei@tencent.com}

\author{Yong Xu}
\affiliation{%
  \institution{Tencent Inc.}
  \city{Beijing}
  \country{China}}
  \email{rogerxu@tencent.com}

\author{Juhong Wang}
\affiliation{%
  \institution{Tencent Inc.}
  \city{Beijing}
  \country{China}}
\email{julietwang@tencent.com}

%\author{Meng Chen$^1$, Hongyu Zhang$^2$, Chengcheng Wan$^3$, Zhao Wei$^4$, Yong Xu$^4$, Juhong Wang$^4$, Xiaodong Gu$^{1}$ %\\
%  $^1$School of Software, Shanghai Jiao Tong University, China\\
%  $^2$Chongqing University, China 
%  $^3$East China Normal University
%  $^4$Tencent Inc\\
%  \{chenmeng,xiaodong.gu\}@sjtu.edu.cn, hongyu.zhang@cqu.edu.au, ccwan@ecnu.edu.cn\\
%  \{zachwei, rogerxu, julietwang\}@tencent.com\\
%}

\pagestyle{plain}

\begin{abstract}

Large language models (LLMs) such as ChatGPT have shown remarkable capabilities in code generation. \rev{Despite significant achievements, they rely on enormous training data to acquire a broad spectrum of open-domain knowledge. Besides, their evaluation revolves around open-domain benchmarks like HumanEval, which primarily consist of programming contests. Therefore, it is hard to fully characterize the intricacies and challenges associated with particular domains (e.g., web, game, and math).}
In this paper, we conduct an \rev{in-depth study of the LLMs in domain-specific code generation}. \rev{Our results} demonstrate that LLMs exhibit sub-optimal performance in generating domain-specific code, due to their limited proficiency in utilizing domain-specific libraries. 
We further observe that incorporating API knowledge as prompts can empower LLMs to generate more professional code.  
%\hy{based on the current results, perhaps this paper can focus on an experimental comparison of various domain-specific code generation models, including DomCoder-kgGPT, DomCoder-CoT (prompting), DomCoder-CoT (fine tuning), DomCoder with/without API, etc., instead of proposing a new code generation model. Can also say that the results can be further improved also suggest possible future works for improving the results. This requires some changes in paper structure though. The current paper title need not change.}
Based on these findings, we further investigate how to effectively incorporate API knowledge into the code generation process. We experiment with three strategies for incorporating domain knowledge, namely, external knowledge inquirer, chain-of-thought prompting, and chain-of-thought fine-tuning. 
We refer to these strategies as a new code generation approach called \textit{DomCoder}. Experimental results show that all strategies of DomCoder improve the effectiveness of domain-specific code generation under certain settings. %The results also show that there is still ample room for further improvement, based on which we suggest possible future works.
\end{abstract}

%\keywords{large language models, code generation, domain-specific program generation}

\maketitle

\section{Introduction}

Large language models (LLMs), such as ChatGPT, have demonstrated remarkable proficiency in various coding tasks~\cite{nascimento2023artificial,hansson2023code}, including code generation~\cite{chen2021codex,nijkamp2022codegen}, software testing~\cite{siddiq2023exploring}, program repair~\cite{cao2023study}, \rev{and code refinement~\cite{guo2024exploring,depalma2024exploring}}. 

Despite their remarkable performance, current LLMs for code generation heavily rely on enormous training data to acquire a broad spectrum of open-domain knowledge. Notably, the evaluation of present LLMs typically revolves around open-domain benchmarks like HumanEval \cite{chen2021codex} and MBPP~\cite{austin2021mbpp}, which primarily consist of programming contests (e.g., sorting, dynamic programming), and while they showcase the capabilities of LLMs in certain aspects, they do not fully represent the intricacies and challenges associated with real-world code generation scenarios\rev{~\cite{liu2024your,jin2024can}}.

In this paper, we explore a more demanding code generation scenario, focusing on domain-specific code generation.  
\rev{We define \emph{domain-specific code} as source code that is tailored specifically for and can only be applied to a particular domain (e.g., web and game), typically developed using domain-specific frameworks (e.g., HTTP, RPC, Unreal).
Unlike general-purpose code, domain-specific code presents distinct challenges due to the scarce availability of code corpora tailored to a specific domain.} \rev{This scarcity is not just about the limited amount of training code available; rather, it highlights the relative rarity of domain-specific training data compared to the extensive pretraining data that spans multiple domains. As a result, LLMs may be short of expertise in these specific domains.}  \rev{This scarcity presents a significant obstacle to the task of generating domain-specific code.}

We aim to answer the following major questions:
(1) How effective are LLMs such as ChatGPT on domain-specific code generation? 
(2) How to effectively prompt LLMs to produce domain-specific code? 
In addition, noticed that it is not always straightforward to guide LLMs on domain-specific code repositories, we also explore the following research question: 
(3) Can we enhance code generation models for a particular domain? Specifically, how to efficiently integrate domain knowledge into code generation models to enable them to excel in domain-specific code generation tasks?

To answer these questions, we constructed a domain-specific code dataset that involves two distinct domains and two programming languages: web development in Go and game development in C++. To ensure that the code belonged to the domain we were interested in, we specifically focus on six prominent industry libraries within the chosen domains, namely, Gin, Prometheus, gRPC-go, Unreal Engine, Cocos2d-x, and Bgfx.  
We curate relevant repositories from GitHub, employing a filtering process that selects code functions utilizing the aforementioned domain libraries. 
%The dataset covers two domains and two programming languages: web development in Go and game development in C++. Each domain includes three libraries, and there are 64,480 functions per library on average. 

We first investigate the abilities of several LLMs (ChatGPT~\cite{ChatGPT}, PolyCoder~\cite{xu2022polycoder}, and CodeLlama~\cite{roziere2023codellama}) in specific domains by comparing their performance on general-purpose code corpora and domain-specific code corpora. 
Our analysis reveals that although LLMs have made remarkable advancements in generating code for open-domain applications, their performance degrades sharply when applied to specific domains.
For ChatGPT, %the BLEU score drops by 70.35\% on average and 
the CodeBLEU score drops by 51.48\% on average. 
We particularly notice that this is often caused by a lack of domain knowledge, particularly the misuse of third-party libraries.   

We then investigate how to effectively prompt LLMs using domain knowledge. We design several basic knowledge-based prompts and use them to elicit ChatGPT, the most popular LLM for code generation. We experimented with different combinations of the basic prompts to study how different types of prompts influence the performance of domain-specific code generation. 
We observe result improvement consistently on all library-specific datasets when using knowledge-enhanced prompts such as API sequences and docstrings compared to using the plain function signature prompt.

Based on these findings, we propose a new method called \ourmethod for domain-specific code generation. \ourmethod integrates domain knowledge into the code generation process of LLMs through three strategies, including 1) inquiring an external knowledge GPT (denoted as kg-GPT), 2) Chain-of-though prompting (CoT-PT), and 3) Chain-of-thought fine-tuning (CoT-FT). 
Kg-GPT trains an external API enquirer based on GPT. Given a function signature, it predicts the API call sequence and uses it to prompt LLM to complete the function body. 
%We utilize the chain-of-thought (CoT) technique~\cite{Chain-of-Thought}, which has proven effective in handling complex multi-step generation tasks across multiple domains including software engineering \cite{}. 
In the chain-of-thought strategies, we conceptualize a domain function as a series of sequential states, where each state corresponds to a sub-activity related to an API call to a third-party library. 
We simulate a chain-of-thought process during code generation: at each step, the model generates a `knowledge state' suggesting relevant APIs and a `task state' describing the intended action to be performed. A knowledge state is predicted based on the history of preceding steps, acting as a guide for the code generation at the current step and augmenting domain-specific knowledge. The process continues until the entire programming task is completed. 
We implement two distinct variants of the chain-of-thought strategies, encompassing both the zero-shot and fine-tuning paradigms. Concerning zero-shot code generation (CoT-PT), we sequentially predict APIs using kg-GPT and leverage each API prediction to prompt subsequent states in the sequence. 
In the fine-tuning setting (CoT-FT), we enrich each of the primary training functions by injecting pre-processed knowledge states. These enriched functions are then used to fine-tune LLMs in an end-to-end fashion.

We apply these strategies on the state-of-the-art code LLMs such as PolyCoder \cite{xu2022polycoder}, StarCoder \cite{li2023starcoder}, and CodeLlama \cite{roziere2023codellama} under both zero-shot and fine-tuning paradigms. 
Our experimental results show that all strategies of \ourmethod improve the effectiveness of domain-specific code generation under certain settings. %Notably, our method exhibits an average improvement of 17.10\% in BLEU scores and 4.20\% in CodeBLEU scores.
%The results also show that there is still ample room for further improvement, based on which we suggest possible future works. 

%The rest of the paper is organized as follows. In Section 2, we provide an overview of related work in the area of large language models for code generation. In Section 3, we describe the empirical study we conducted. In Section 4, we describe our approach and the methodology used in our experiments. In Section 5, we present the experimental results and analyze the performance of our models. Finally, we conclude the paper in Section 6 and discuss future research directions.

The major contributions of this paper are summarized as follows:
\begin{itemize}
    \item An empirical study on the ability of LLMs for domain-specific code generation. % across two domains, involving 6 domain libraries.
   \item An investigation of the effectiveness of various prompts for LLMs to generate domain-specific code.
   \item A new approach to integrate domain knowledge into LLMs for code generation.
   \item A thorough discussion about the implication of our findings and future research directions.
\end{itemize}

\section{Related Work}

%VisProg~\cite{gupta2023visprog}

%\gu{check out these papers: https://arxiv.org/pdf/2107.07112.pdf, https://arxiv.org/pdf/2207.05579.pdf,
%https://www.ijcai.org/proceedings/2021/0512.pdf
%https://aclanthology.org/2021.emnlp-main.482.pdf
%}

\subsection{Code Generation}

Code generation aims to automatically synthesize programs conditioned on programming language context and/or natural language descriptions \cite{chen2021codex,xu2022polycoder,black2022gptneox}, therefore increasing productivity for developers. Recently, LLMs for code generation have gained significant popularity due to their impressive performance. The state-of-art models are based on GPT, due to their auto-regressive nature, which is suitable for generative tasks where the model predicts the next tokens given the previous context.

General-purpose LLMs exhibit strong code-generation capabilities. %, though not specifically trained on code corpus. 
These models include GPT-Neo~\cite{gpt-neo}, GPT-J~\cite{gpt-j} and GPT-NeoX~\cite{black2022gptneox}. 
One representative model is OpenAI's ChatGPT~\cite{ChatGPT}, a model designed for conversation tasks and supported by a family of backbone models including gpt-3.5-turbo and GPT-4. It is trained on WebText~\cite{Radford2019LanguageMA}, OpenAI's internal corpus collected from web pages. 

There are also many LLMs designed for code-related tasks that are trained on large publicly available code corpora such as the GitHub repositories.
%The large data size allows them to understand and generate programming languages effectively. %These models excel especially in solving general coding problems such as language comprehension, algorithms, and simple mathematics. 
For example, Codex~\cite{chen2021codex} is a GPT language model aimed at synthesizing Python programs from docstrings. %and performs well on the HumanEval~\cite{chen2021codex} benchmark. 
CodeParrot~\cite{tunstall2022codeparrot}, CodeGen~\cite{nijkamp2022codegen}, CodeGeeX~\cite{zheng2023codegeex}, PolyCoder~\cite{xu2022polycoder}, StarCoder~\cite{li2023starcoder}, \rev{and CodeLlama}~\cite{roziere2023codellama} also belong to this category.
Compared to our works, they mainly focus on general-purpose code generation rather than specific domains.

% For example, Codex~\cite{chen2021codex} is trained on the Pile~\cite{gao2020pile}, a corpus containing text and code from a variety of sources. Similarly, 
% CodeGen~\cite{nijkamp2022codegen} is trained on a multi-lingual dataset named BIGQUERY \cite{}, and PolyCoder~\cite{xu2022polycoder} is trained on GitHub code. These training datasets are collected without regard to domains. Furthermore, the evaluation benchmark HumanEval~\cite{chen2021codex} also contains only general-purpose coding problems as well. %such as language comprehension, reasoning, algorithms, and simple mathematics.
% Comparatively, our work focuses on improving the performance of code generation by only using the domain-specific code dataset to construct prompts or as fine-tuning data}.

\subsection{Prompting Large Language Models}

With the capability of natural language understanding and generating, LLMs can be prompted to perform a wide range of downstream tasks, such as question answering~\cite{chen2023gotta}, security~\cite{zhao2023prompt}, and programming~\cite{wang2022prompt-tuning-code,feng2023prompting}. 
Like fine-tuning, prompting is also a means of transferring pre-trained models to a certain downstream task~\cite{kojima2022large}. When it comes to large models such as GPT-4, prompting is often preferred as it does not require a large amount of training resources.

Prompts can be broadly classified into two categories: soft prompts and hard prompts.  
Soft prompts are continuous vectors that can be optimized by training. 
For example, Li and Liang~\cite{prefix-finetune} and Lester et al.~\cite{power-of-scale} explored prompt tuning for learning soft prompts.  
Hard prompts are interpretable text tokens that describe the task and may include examplars.
For example, Reif et al.~\cite{ReifIYCCW22} augmented discrete text prompts to provide more detailed descriptions.
Recently, Chain-of-Thought (CoT) Prompting~\cite{Chain-of-Thought} has emerged as a prevalent way to engage LLMs on specific tasks. CoT prompting has leveraged exemplars with intermediate steps to enhance the reasoning capabilities of LLMs in complex reasoning tasks.
% enhances the reasoning capabilities of large language models by breaking down multi-step problems into intermediate steps and using them as prompting exemplars, allowing language models to tackle complex reasoning tasks that cannot be solved with standard prompting techniques. 

The use of prompting techniques has also been extended to code generation~\cite{wang2022prompt-tuning-code,liu2023improving,choi2023codeprompt,cheng2023prompt}. Some researchers optimize the prompting patterns. Liu et al.~\cite{liu2023improving} and White et al.~\cite{white2023chatgpt} investigated various prompt templates for ChatGPT. Some works enhance prompting by incorporating contextual information. For example, Liu et al.~\cite{liu2023improving} added to the prompt optimization template extra context such as Java Class information. Li et al.~\cite{abs-2303-17780} introduced test cases and API information to retrieval-based prompts.

While previous studies have used a function as the unit of prompting, our study focuses on more fine-grained knowledge integration strategies by injecting chain-of-thought prompts into substeps during the code generation process.

\subsection{Code Generation with Domain Libraries}

Recently, code generation utilizing domain libraries has attracted research attention and there have been numerous works focused on incorporating knowledge into code generation models \cite{zan2022language,2023liucodegen4libs}. For example, Zan et al. \cite{zan2022language} %address the challenge of incorporating knowledge about private APIs not typically exposed during model training. 
propose a retrieval-then-coding framework where an API retriever first identifies useful APIs and an API Coder generates code using these APIs. 
Shrivastava et al. \cite{shrivastava2023repository} introduce domain-specific knowledge in the prompt design process. They present the Repo-Level Prompt Generator framework, which generates example-specific prompts for LLMs. This framework incorporates context from entire repositories.
Liu et al. \cite{2023liucodegen4libs} introduce CodeGen4Libs, a technique specifically designed for library-oriented code generation. %Their technique involves two stages: import generation and code generation. By considering the given natural language query and third-party libraries, 
CodeGen4Libs first generates import statements and then concrete code.
%Other works utilize sketch-based approaches to enhance the code-generation process. 
Zan et al. \cite{zan2022cert} leverage the fact that library-oriented code snippets often share similar code sketches. They propose a strategy involving a sketcher and a generator trained on unlabelled data. The sketcher generates library-oriented code sketches, then the generator predicts code snippets based on them.

Our work differentiates itself from existing studies by exploring deeper integration of domain (API) knowledge into the code-generation process. For example, we propose a GPT-based API recommender that generates API knowledge prompts for LLMs. We also formulate code generation as a chain-of-thought process and integrate APIs in each internal coding state. Besides, we provide an in-depth analysis of the knowledge gap of LLMs in generating domain code.

%The semantics and syntax of programming language differentiate the generation and evaluation of code from those of natural language. 
%In code, the same functionality can be implemented in different manners, resulting in low token-level similarity even though they are semantically equivalent. Therefore simple sequence similarity is not sufficient for code evaluation. 
%CodeBLEU\cite{} injects code syntax evaluation via abstract syntax trees (AST) and code semantics evaluation via data flow, enhancing BLEU for a comprehensive comparison of generated code and the ground truth.

%The pass rate on test cases has also been utilized as a benchmark for code generation evaluation, such as HumanEval.

\subsection{Studies on Code Generation with LLMs}
\rev{
In addition to our research, numerous empirical studies have explored code generation with LLMs~\cite{mastropaolo2023robustness,liu2024no,liu2023your,vaithilingam2022expectation}. For instance, Vaithilingam et al.~\cite{vaithilingam2022expectation} conducted a user study to understand how programmers use and perceive LLM code generation tools. They found that while Copilot reduced the need for online searches, it introduced new challenges in code understanding, editing, and debugging.
Mastropaolo et al.~\cite{mastropaolo2023robustness} evaluated the robustness of existing code-generation LLMs, discovering that developers using different wordings to describe the same code received varying recommendations.
Liu et al.~\cite{liu2023your} hypothesized that existing benchmarks, which rely on curated synthesis problems and test cases, may be insufficient for fully assessing the functional correctness of generated code. They proposed EvalPlus, a code synthesis evaluation framework designed to rigorously benchmark the functional correctness of LLM-generated code .
More recently, Liu et al.~\cite{liu2024no} conducted a systematic empirical assessment of the quality of code generation using ChatGPT, revealing potential issues and limitations in ChatGPT-based code generation.
}

\rev{
Our research stands out from previous studies in both its scope and methodologies. Whereas many related studies focus on evaluating the quality and usability of generated code and challenge existing benchmarks, our work emphasizes a crucial aspect of code language models: their proficiency in generating domain-specific code. Specifically, we concentrate on how well LLMs acquire and utilize domain-specific libraries and their overall effectiveness in applying domain knowledge.
}

\section{Experimental Evaluation}

%Large language models (LLMs) achieve remarkable results in general code-generation tasks. However, their code-generation capability for specific domains has not been extensively studied, despite its significance to developers. 
%LLMs are typically trained with code snippets from the general application domain such as all the public repositories hosted on GitHub, where the domain-specific code only occupies a small portion. It brings up a threat of limited code generation capability in these domains. 
%To address this gap, 
In this paper, we conduct an empirical study of LLMs on domain-specific code generation. 
%In this section, we describe our empirical study on domain-specific code generation by large language models.
Specifically, we aim to address the following research questions:

\textbf{RQ1}: How effective are LLMs in domain-specific code generation?

\textbf{RQ2}: What are effective methods for prompting LLMs to produce domain-specific code?

%\textbf{RQ3}: How to efficiently integrate domain knowledge in the code generation process? 

 \begin{table*}[htbp]  
   % \small
    \centering
    \caption{Statistics of domain-specific libraries}
    \scalebox{0.9}{
    \begin{tabular}{lccccccc}  
        \toprule
        \bf Library &\bf Language &\bf Domain &\bf Description  &\bf Popularity &\bf \# train functions \\
        \midrule
        Gin & Go & web development & web framework featuring Martini-like APIs& 69.5k stars & 21,161 \\
        gRPC-go & Go & web development & remote procedure call across data centers & 18.3k stars & 129,803 \\
        Prometheus & Go & web development &  client library for application instrumentation& 9.1k stars & 11,708 \\
        Unreal Engine &  C++ & game development  &  realtime 3D game engine& - & 184,808 \\
        Cocos2d-x & C++ & game development &  cross-platform 2D game framework & 17.2k stars & 27,119 \\
        Bgfx & C++ & game development  & cross-platform rendering library & 13.1k stars & 12,285 \\
        \bottomrule
    \end{tabular} 
    }
\label{libs}
\end{table*} 

\subsection{Data Collection}
\label{sec:data}
To evaluate LLMs on domain-specific code generation, we collect data from the public repositories on GitHub that involve certain domains. As the domain is not usually explicitly labeled in GitHub repositories, we track the domain of each function through the third-party libraries they utilize. 
We consider two popular domains in our experiments, including web and game development. The third-party libraries of these two domains are clear for identification. 
% library details
For each domain, we investigate three libraries provided by a giant internet company, voted as the most commonly used in their domain applications. 
Table~\ref{libs} shows the details of all libraries, including the languages, domains, descriptions, and popularity represented by GitHub stargazer counts.
For web development, we chose three libraries in the Go language, including Gin, gRPC-go, and Prometheus.
\emph{Gin} (\texttt{gin-gonic/gin}) is a web framework that features a Martini-like API with better performance.
\emph{gRPC-go} (\texttt{google.golang.org/grpc}) is an implementation of gRPC, a high-performance, general-purpose, open-source RPC framework designed by Google.
\emph{Prometheus} (\texttt{prometheus/client\underline{~}golang}) is a client library for instrumenting Go applications. It has two separate parts, one for instrumenting application code, and one for creating clients that talk to the Prometheus HTTP API.
%\emph{Protobuf} (google.golang.org/protobuf ) contains Go bindings for protocol buffers, a language-agnostic binary data format suitable for network communications.

For game development, we chose three C++ libraries: \texttt{Unreal Engine}, \texttt{cocos2d-x}, and \texttt{bgfx}. 
\emph{Unreal Engine} is a real-time 3D game development engine widely used in the industry. With solid network support, it can be used to create multi-player online games that require high performance and low latency.
\emph{Cocos2d-x} is an open-source multi-platform C++ framework for building graphical applications, especially 2D games. It works on platforms including iOS, Android, macOS, Windows, and Linux.
\emph{Bgfx} is an open-source cross-platform rendering library,  that can be used with graphic libraries to build game frameworks.

%Our final dataset contains code for six libraries covering two languages, Go and C++, and two domains, web development, and game development. Considering the practicality of our research,  the libraries were chosen from the commonly used ones in the industry.

 %\begin{table}[htbp]  
 %  % \small
 %   \centering
 %   \caption{Data statistics}
 %   % \setlength{\tabcolsep}{1mm}{
 %   \scalebox{1.0}{
 %   \begin{tabular}{lccc}  
 %       \toprule
 %       \bf Library &\bf \# training functions  \\
 %       \midrule
 %       %\hline
 %       gin  & 21,161 \\
 %       grpc-go & 129,803 \\
 %       prometheus & 11,708 \\
 %       Unreal Engine & 184,808 \\
 %       cocos2d-x & 27,119 \\
 %       bgfx & 12,285 \\
 %       \bottomrule
 %   \end{tabular}
 %   }
 %   % }
 %   \label{dataset}
%\end{table}    

% collection
The detailed data collection process is as follows. We first filter all publicly available repositories with above 50 Stargazer counts by the target language (e.g., \texttt{Golang}) using the GitHub repository search API\footnote{https://docs.github.com/en/rest/search?apiVersion=2022-11-28}. Then, for each repository filtered, we use the GitHub code search API to check if it contains the library name which must be imported, and if so, retrieve all program files that contain the library name. 
We extracted and deduplicated all functions from the collected files, then filtered out those with an empty function body.
We also leave out functions with simple names such as \textit{main} or \textit{init}, as they do not contain sufficient context information about the functionality of the code. 
The statistics of our final dataset are shown in Table~\ref{libs}.

\subsection{Evaluated Language Models}
%We conduct the empirical study on three models, ChatGPT, CodeGen, and PolyCoder. To make our results representative, we chose large models that have been recently released and widely used, including models for general text generation and models for code generation. An overview of the model settings is presented in table\ref{models}.
%As summarized in Table~\ref{models}, 

We investigate three LLMs: ChatGPT \cite{ChatGPT}, CodeLlama~\cite{roziere2023codellama}, and PolyCoder~\cite{xu2022polycoder}. They are all widely used for code generation.

%  \begin{table}[htbp]  
%     \centering  
%     \caption{Evaluated models\hy{can omit if space is limited}}
%     % \setlength{\tabcolsep}{1mm}{
%     \scalebox{1.0}{
%         \begin{tabular}{lcccccc}  
%             \toprule
%             \bf Model &\bf Version &\bf Size &\bf Release Date \\ %&\bf Task \\
%             \midrule
%             \ ChatGPT & gpt-3.5-turbo  &  154B    &  2022   \\ %&  Text Generation   \\
%             \ CodeGen & CodeGen-2B-multi &  2B    &  2022  \\ % &  Code Generation     \\
%             \ PolyCoder & PolyCoder-2.7B &  2.7B    &  2022  \\ % &  Code Generation   \\
%             \bottomrule
%         \end{tabular}
%         }
%     % }
%     \label{models}
% \end{table}    

\textbf{ChatGPT}~\cite{ChatGPT} is a popular LLM that performs generation tasks in the dialogue format. The model also has demonstrated superb ability in coding. ChatGPT involves a family of backbone models. We chose the most widely-used version \texttt{gpt-3.5-turbo} to experiment with. The model has around 154B parameters. Since it is not open-sourced, we connect to the model through the official API provided by OpenAI\footnote{ChatGPT: https://platform.openai.com/docs/api-reference} once at a time, then retrieve the code segment from its response text. %\wan{I feel that these three paragraphs are a bit redundant with section 2. If we want to keep them, probably we could directly say which version/setting we are using}

%\textbf{CodeGen}~\cite{nijkamp2022codegen} is a family of open-source models for program synthesis. We chose the version CodeGen-Multi 2B, which is pre-trained on a multi-lingual dataset BIGQUERY~\cite{BigQuery} (C, C++, Go, Java, JavaScript, and Python) and has a size of 2B. We use the released pre-trained checkpoints\footnote{CodeGen: https://huggingface.co/Salesforce/codegen-2B-multi} to conduct generation on each prompt.

\textbf{PolyCoder}~\cite{xu2022polycoder}:
PolyCoder is a well-established open-source model based on GPT-2 and trained on GitHub code. The model demonstrates competitive performance to GPT-3 counterparts. We use the released pre-trained checkpoint PolyCoder-7B\footnote{PolyCoder: https://huggingface.co/NinedayWang/PolyCoder-2.7B}.

\textbf{CodeLlama}~\cite{roziere2023codellama} is the state-of-the-art language model for code generation. CodeLlama is built upon Llama 2, by further training on code-specific datasets. The model involves four sizes with 7B, 13B, 34B, and 70B parameters respectively. We use the 7B model, which can be served on a single GPU while showing efficient and accurate performance on code completion.

For each model, we set the maximum generation length to 256. % \rev{, in alignment with the maximum code length in HumanEval, which makes the generation fast while being capable of handling the main part of code samples}.

\subsection{Evaluation Metrics}

We did not adopt the widely used \textbf{pass@k} (the passing rate of generated code on test cases) \cite{chen2021codex} for evaluation. The rationale behind this choice is rooted in the distinctive nature of the domain-specific functions we investigate. Unlike programming contests in existing benchmarks~\cite{chen2021codex,nijkamp2022codegen}, domain-specific functions tend to be notably intricate, often interrelated with multiple functions, designed for specific platforms, or reliant on a lot of third-party libraries. It is too rigorous to have all of them executable, even with powerful LLMs. For example, our preliminary experiments indicate that even ChatGPT achieves a 0\% pass rate out of 200 samples when attempting to pass their test cases. 
Furthermore, testing them necessitates the labor-intensive task of manually crafting hundreds of test cases, thoroughly constructing multiple functions, and configuring reliant libraries and platforms, which is an impractical endeavor. %To illustrate this, consider the following example, in which PolyCoder-generated code encountered build error, due to several non-existent APIs.
%\gu{Provide an illustrative example showcasing the impracticality of running tests on a domain-specific function.}
%\begin{figure}[htbp]
%    \centering
%    \fbox{
%        \includegraphics[width=0.8\columnwidth, trim=0 0 0 0 clip]{figures/error_running_tests.pdf}
%    }
%    \caption{An instance of non-executable code generated by PolyCoder}
%    \label{fig:eval_metrics}
%\end{figure} 
Given these inherent complexities and limitations, conducting evaluations based on real executions is infeasible.

Therefore, in our study, we have chosen to utilize other popular metrics such as BLEU and CodeBLEU, \rev{which have also been employed by related works on code generation~\cite{li2023skcoder,2023liucodegen4libs}}.
BLEU~\cite{papineni2002bleu} measures sequence similarity by calculating their n-gram matches. 
CodeBLEU~\cite{ren2020codebleu} is a metric based on BLEU, adjusted for code evaluation, which introduces syntactic and semantic similarities into comparison.
We compute both metrics using the scripts provided by the CodeXGLUE paper\footnote{https://github.com/microsoft/CodeXGLUE/tree/main/Code-Code/code-to-code-trans/evaluator/CodeBLEU}. 

%\textbf{Perplexity (PPL)}\cite{} is a common metric for evaluating causal language models, it is calculated as an exponent of the loss obtained from the model.

 \begin{table*}[htbp] 
    \centering
    \caption{The code generation performance of LLMs on open-domain and domain-specific datasets}
%\gu{gRPC shows sharp differences between gpt-3.5-turbo and the other two LLMs. Can you double-check the results?}\cm{ChatGPT tends to give answers like "//TODO: implement the function" for grpc-go}\hy{perhaps can also add Role Playing in the prompts, such as You are a programming assistant; You are a programming assistant for xx domain, etc}}
    % \setlength{\tabcolsep}{1mm}{
        \begin{tabular}{lccccccc}  
            \toprule
              &   \multicolumn{2}{c}{\it\bf General-Purpose} & & \multicolumn{4}{c}{\it\bf Code-Oriented} \\ \cline{2-3}\cline{5-8}
            \multirow{2}{*}{\bf Dataset}  & \multicolumn{2}{c}{\it ChatGPT-3.5-154B} & & \multicolumn{2}{c}{ \it \rev{CodeLlama-7B}} & \multicolumn{2}{c}{\it PolyCoder-2.7B} \\
            &\bf BLEU &\bf CodeBLEU & & \bf  BLEU & \bf CodeBLEU & \bf BLEU & \bf CodeBLEU  \\
            \midrule
            %\ Paper  &  &  & &  - & - & &  \\
            %\hline
            \bf Open-domain datasets & \\
             \ HumanEval & 39.00 & 30.05 & & 18.13 & 17.99 & 22.22  & 13.81  \\
            %\ CodeSearchNet (Golang) & \bf10.59 & \bf20.86 & &\bf13.72 & \bf17.63 & \bf14.95  & \bf18.39  \\
             % resampled
             \ CodeSearchNet (Golang) & 11.16 & 25.29 & & \rev{16.14} & \rev{19.63} & 15.97  & 19.56    \\
             %\bf \ \rev{Average} & \rev{25.08} & \rev{27.67}  &  & \rev{17.14}  &  &   &   \\
             \hline
             \bf Domain-specific datasets & \\
             \ Gin & 5.81 & 17.73 & & \rev{8.47} & \rev{18.07} & 8.13 &  17.05 \\
             \ Prometheus & 1.75 & 12.17  & & \rev{1.79} & \rev{13.13} & 1.77 & 12.11   \\
             %\ Protobuf & 4.80  & 10.53  &  &  &  & 12.29 & 17.23      \\
             \ gRPC-go & 2.53  & 12.15  &  & \rev{57.61} & \rev{60.39} & 55.36  & 57.52    \\
             \ Unreal Engine & 5.22  & 10.57  & & \rev{0.73} & \rev{10.21} & 0.94  &   9.87  \\
             \ cocos2d-x   &  3.71 & 12.92  & & \rev{16.76} & \rev{25.87} & 13.83  &  23.83   \\
             \ bgfx   & 0.86  & 8.09 &  & \rev{0.55} & \rev{7.72} &  0.76 &  7.16   \\
             %\bf \ \rev{Average} & \rev{3.31} &  &  & \rev{14.32} & \rev{22.57} & \rev{13.46} & \rev{21.25} \\
            \bottomrule
    \end{tabular}
    % }
    \label{table_rq1}
\end{table*}    

\subsection{RQ1: Effectiveness of LLMs in domain-specific code generation}
%In this section, we present both quantitative results and manual evaluation results for a comprehensive analysis. 

\subsubsection{Methodology}

In this RQ, we investigate the capabilities of LLMs in specific domains. We focus on the code completion scenario, wherein an LLM is presented with a function signature as a prompt, and its objective is to complete the function body. \rev{Previous work has indicated that a good function signature already provides enough information to illustrate its functionality} \cite{dingcode}. \rev{Another reason for not including NL requirements in the prompt is that the domain-specific code we collected contains only a small portion ($<$5\%) of code that contains NL requirements.} 

%We investigate domain-specific code generation on the web-development datasets collected from GitHub, which contains code utilizing each of the three Go libraries (Gin, Prometheus, gRPC-go). 

%For each library, we collected a dataset containing functions using that library. 

We sampled 500 functions from each library-specific dataset we collected (Section~\ref{sec:data}) for the completion. We compare the model-generated code against the ground-truth one in the original function and compute the evaluation metrics. %We calculate PPLs based on the probabilities predicted by each LLM. 
% We also take the perplexity of PolyCoder calculated on open-domain validation data into comparison.

We compare the quality of LLM-generated code in the domain-specific dataset to that in open-domain (general-purpose) datasets and then analyze the performance gap.
We use HumanEval and CodeSearchNet (Golang)~\cite{husain2019codesearchnet} as the open-domain code corpora. HumanEval is a widely-used benchmark for evaluating code generation models~\cite{chen2021codex,nijkamp2022codegen}. CodeSearchNet is a benchmark for evaluating code comprehension tasks. The benchmark was collected from miscellaneous projects in GitHub and was processed unbiasedly.

\subsubsection{Quantitative Results}

Table~\ref{table_rq1} shows the quantitative results. Overall, all three models demonstrate a decline in performance on domain-specific datasets compared to open-domain datasets, which conforms to our hypothesis. 
For ChatGPT-3.5, the average BLEU score on domain-specific datasets drops by 70.35\% compared to the score on CodeSearchNet, the average CodeBLEU drops by 51.48\%.
% Similarly, the code-oriented models perform worse on most of the domain-specific datasets. 
% On average, The BLEU score drops by 61.10\% for ChatGPT-3.5, 57.51\% for CodeGen-2B-multi, and 50.50\% for PolyCoder-2.7B. The CodeBLEU score drops by 49.81\% for ChatGPT-3.5,  16.90\% for CodeGen-2B-multi, and 15.93\% for PolyCoder-2.7B.

We have noted that \rev{non-popular libraries} (such as Prometheus) tend to experience a more significant performance decline compared to others. This is likely due to the fact that there is less open-source code that utilizes the domain-specific library available for LLM training. On the contrary, gRPC-go and cocos2d-x exhibit exceptionally high performance in code-oriented LLMs. 
%This notable performance can be attributed to the extensive prevalence of code utilizing these libraries in the training set of coding LLMs. %\wan{There is no much differences in their Github stars/forks. Can we count the number of occurance of that uses these frameworks in the training data? Then we could have some concrete numbers.} 
Upon manual examination, we observe monotonous code patterns in functions that utilize these two libraries, such as gRPC server handler functions and cocos2d-x interface functions from C++ to Lua. Such repetitive knowledge is more easily learned by the models.

Notably, ChatGPT outperforms the code-oriented LLMs in the open-domain dataset. 
This phenomenon can be explained by the fact that general-purpose LLMs are trained on a vast amount of diverse data, which enables them to learn a broad range of knowledge. This makes them better suited for open-domain tasks requiring a more general language understanding.
%On the other hand, code-oriented LLMs are trained explicitly on code corpus, allowing them to better understand programming concepts and be more effective than general-purpose LLMs in domain-specific datasets.\hy{then why the CodeBLEU scores are similar?}

\begin{figure}[htbp]
    \centering
    \subcaptionbox{Misused API calls}{
        \fbox{
            \includegraphics[width=0.9\columnwidth, trim=0 0 0 0 clip]{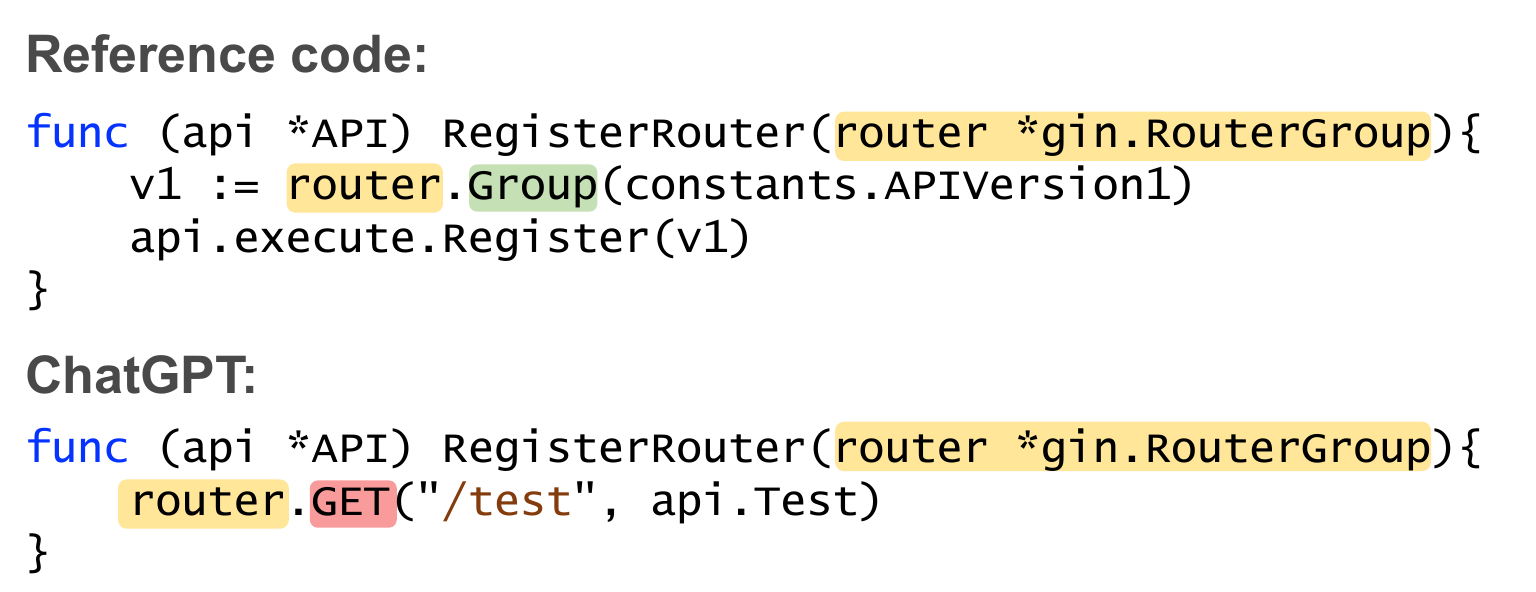}
            % TOSEM    Conference:0.9
        }
        \label{fig:rq1:eg1}
    }
    \subcaptionbox{Missing API calls}{
        \label{fig:rq1:eg2}
        \fbox{
        \includegraphics[width=0.9\columnwidth, trim=8 0 5 0 clip]{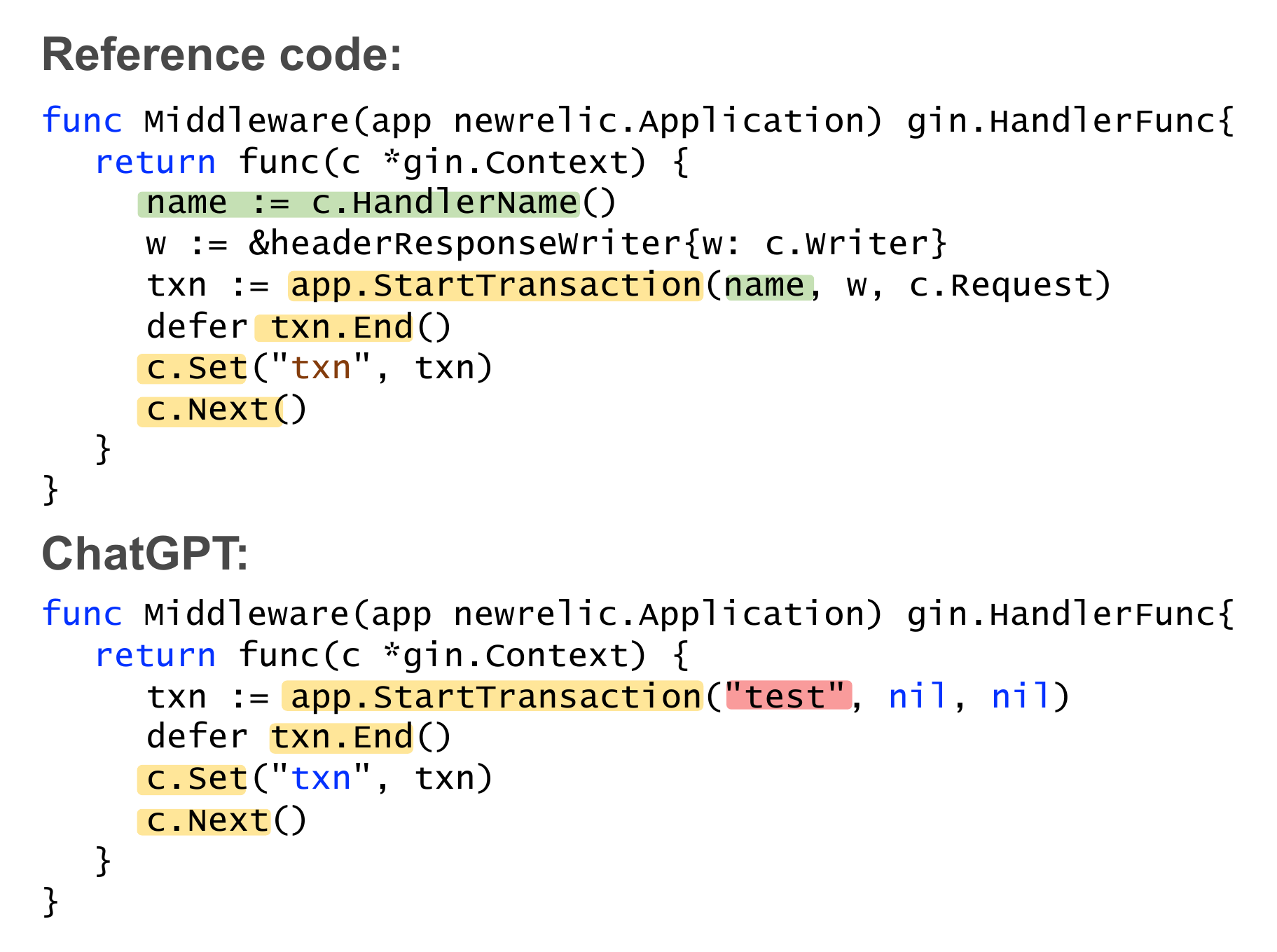}
        }
    }
    \caption{Code examples generated by ChatGPT}
    \label{fig:rq1:eg}
\end{figure}

 \begin{table*}[htbp]  
   % \small
    \centering
    \caption{Examples of knowledge-enhanced prompts}
    \scalebox{0.95}{
        \begin{tabular}{ll}  
            \toprule
             \bf Prompt type & \bf Example \\
            \midrule
            % \hline
            \bf basic prompts & \\
             \ Function signature & \small \texttt{Complete this function:} \textcolor{cyan}{\texttt{func Routes(r *gin.Engine)}}\\
             \ Library import & \small \texttt{Complete this function \textcolor{cyan}{using gin}: func Routes(r *gin.Engine)} \\ 
             \ API & \small \texttt{Complete this function \textcolor{cyan}{using gin.RouterGroup.Use}: func Routes(r *gin.Engine)}\\
             \ docstring & \small \texttt{gin.RouterGroup.Use \textcolor{cyan}{adds middleware to the group}.}\\
             & \small \texttt{Complete this function: func Routes(r *gin.Engine)}\\
             \midrule
            \bf combined prompts & \\
             \ API  + docstring & \small \texttt{Complete this function using gin.RouterGroup.Use.}\\
             & \small \texttt{gin.RouterGroup.Use adds middleware to the group.   func Routes(r *gin.Engine)} \\
             \ docstring + API & \small \texttt{gin.RouterGroup.Use adds middleware to the group. }\\
             & \small \texttt{Complete this function using gin.RouterGroup.Use: func Routes(r *gin.Engine)} \\
             
            \bottomrule
        \end{tabular}
        }
        % }
    \label{prompt_examples}
\end{table*}    

\subsubsection{Qualitative Results}

To find out which aspects of the models fall short of domain-specific code generation, we also qualitatively examined the ChatGPT-generated code manually. 
We randomly sample 200 functions from the library-specific dataset. For each function, we use the function signature as a prompt and let ChatGPT complete the function body. 
We check if the generated code has implemented the desired functionality as in the original reference code, and we pay particular attention to the invoke of domain libraries.

Overall, of the 200 model-generated functions sampled, we observed that 32 functions were correct (16\%), 37 implemented functionality completely different from the reference code (18.50\%), 52 had problems with API misuse (26\%), and 79 had problems with missing APIs (39.50\%). 
We hereby summarize common mistakes that appear in the model-generated code (Figure~\ref{fig:rq1:eg}).

1. \textbf{Misused API calls}
The generated code is similar in format to the reference code, but the actual functionality is incorrectly implemented due to wrong API calls, we refer to this kind of mistake as \textit{misused API calls}.
An example is shown in Figure \ref{fig:rq1:eg} (a).
The reference function \textit{RegisterRouter} registers an HTTP API router, with a \textit{gin.RouterGroup} object passed as a parameter.
The correct API to be called is \textit{router.Group}, which returns a router group for later use, while the model-generated code used \textit{router.GET}, which is also an API of the Gin library, but conducts a different activity: handling a GET request. Therefore the generated code functionality is inconsistent with the demand given by the function signature.

2. \textbf{Missing API calls}
In some cases, we observed the model-generated code is missing one or several APIs compared to the reference code, resulting in a deviation of the implemented functionality from the expected functionality. We refer to this issue as missing API Calls.
As shown in Figure~\ref{fig:rq1:eg} (b), the reference code invokes \textit{gin.Con} \textit{text.HandlerName} to retrieve a name for the handler, and passes it as a parameter to \textit{app.StartTransaction}. While in the model-generated code, this API call is missing, leading to wrong parameters for the next API call.

% 3. \textbf{Misplaced API calls}
In summary, most of the mistakes in domain-specific code generation are related to API misusage, leading to our conclusion that existing LLMs are unfamiliar with the API usages of certain domain-specific libraries. %, due to insufficient domain knowledge.

\begin{tcolorbox}[width=\linewidth, boxrule=0pt, sharp corners=all,
 left=2pt, right=2pt, top=2pt, bottom=2pt, colback=gray!20]
\textbf{Finding 1}: LLMs exhibit suboptimal performance in generating domain-specific code, specifically due to their limited proficiency in utilizing domain libraries.
\end{tcolorbox}

\subsection{RQ2: Prompting LLMs for Domain-Specific Code Generation}

Based on the evaluation results in RQ1, we have found that LLMs underperform in the domain-specific code generation task, which we attribute to LLMs' lack of domain-specific knowledge by observation. 
Therefore, we assume that incorporating domain-specific knowledge into LLMs can improve the quality of code generation. % by providing the model with a domain-relevant context.  
In the view of programming, domain knowledge typically refers to the usage of third-party libraries or packages, namely, API documentation. For instance, in web development using Golang, the Gin web framework is commonly used to enhance productivity and performance. The details of the Gin API, such as the names and descriptions of its functions, are examples of domain-specific knowledge.

Another question is how to incorporate domain knowledge into LLMs. As a direct and cost-effective method, prompting has emerged as one of the most popular ways to engage with LLMs~\cite{brown2020gpt3,white2023chatgpt,liu2023improving}. It provides LLMs with a few instructions such as intentions, demonstration examples, and a chain of thoughts, thereby enabling LLMs to produce desired outcomes. We hypothesize that by formulating domain knowledge as prompts, we can effectively induce LLMs to generate domain-specific code more proficiently.

Therefore, in RQ2, we investigate how to effectively prompt LLMs using domain knowledge and how different forms of prompt impact the performance of domain-specific code generation.
To address this, we designed several basic knowledge-based prompts and used them to elicit ChatGPT, the most popular LLM for both text and code generation. ChatGPT also demonstrates a superb ability in understanding human prompts hence it is easier to showcase the effects of different prompts. We study the effect of each kind of knowledge prompt separately. Moreover, we experimented with different combinations of the basic prompts to study how the order of knowledge elements can affect the code generation quality, and how different types of knowledge can complement each other.

 \begin{table*}[htbp]  
    \centering
    \caption{Results of Prompting ChatGPT for Domain-Specific Code Generation (CB denotes CodeBLEU)}
    \scalebox{0.93}{
        \begin{tabular}{lccccccccccccc}  
            \toprule
                     &   \multicolumn{2}{c}{Gin} & \multicolumn{2}{c}{Prometheus} & \multicolumn{2}{c}{gRPC-go} & \multicolumn{2}{c}{Unreal Engine} & \multicolumn{2}{c}{Cocos2d-x} & \multicolumn{2}{c}{bgfx} \\
           \bf Prompt type &\bf BLEU  &\bf CB & \bf BLEU & \bf CB   &\bf BLEU  &\bf CB &\bf BLEU  &\bf CB & \bf BLEU & \bf CB & \bf BLEU & \bf CB \\
            \midrule
            % \hline
             function signature & 5.81 & 17.73 & 1.75 & 12.17 & 2.53 & 12.15 & 5.22 & 10.57 &  3.71 & 12.92 &  0.86 & 8.09 \\
             + library import &  6.08 &  17.83 &  3.37 &  15.12 &   5.44 & 16.93  & 5.79  & 11.27 & 5.37  &  14.68 &  1.02 & 8.79 \\
             + API  & \bf 14.86  & 26.69 &  \bf7.18 & 20.51 & \bf22.24 & \bf31.84 &  \bf14.78 &  18.34 & \bf14.18  & \bf22.35 &  \bf1.39  & 11.32 \\
             ~~~~\,\, + docstring & 14.56 & \bf27.04  & 7.04 & \bf 20.78 &22.07 & 31.74 & 14.58  & \bf18.35 & 14.06 & \bf22.35 & 1.33 &  \bf11.40\\
             + docstring & 5.57 & 17.78 & 1.94 & 12.31 & 2.54 & 12.33  &  5.32 & 10.41 & 3.57 & 12.93 & 0.88 & 8.38\\
             ~~~~\,\, + API  & 13.64 & 25.94 & 6.95 & 20.48 & 22.08 & 31.64 & 14.12  & 17.74  & 14.00 & 22.24 & 1.23 & 11.03 \\
            \bottomrule
        \end{tabular}
        }
        % }
    \label{rq2_result}
    % results for propotus:
    % 4.80, 10.53
    % 5.44， 12.88
    % 9.65， 17.74
    % 10.50， 18.25
    % 6.29， 12.95
    % 10.14， 18.09
\end{table*}    

\subsubsection{Prompt Design}

We regard the function signature as a plain prompt without any knowledge injection. Apart from this, we designed three types of knowledge-based prompts to enhance the code generation quality:

\textbf{1) Library import prompt}: a phrase specifying the third-party library to be used.

\textbf{2) API prompt}: a list of APIs to be called to implement a function,  retrieved from the ground-truth code.

\textbf{3) Docstring prompt}: the natural language descriptions of the APIs,  retrieved from the library documentation.

We also constructed two combined knowledge-based prompts: \textbf{API+docstring} and \textbf{docstring+API.}
To illustrate, we provide examples of each type of knowledge-enhanced prompt for ChatGPT-based code generation in Table \ref{prompt_examples}.

\subsubsection{Results}
Table \ref{rq2_result} shows the performance of ChatGPT on library-specific datasets under various prompt types. We use plain function signatures as a baseline and experiment with each knowledge-enhanced prompt.

We observe result improvement on library-specific datasets when using knowledge-enhanced prompts such as library indication and API name sequence compared to using the plain function signature.
Comparing the results of basic prompts, API sequence demonstrates a significant improvement, whereas library import and docstring exhibit relatively modest enhancements. The API name sequences are closer related to the code content and have a more direct effect on code generation. In contrast, the library import prompts do carry concise and useful information, but in meager quantity. The API docstrings contain implicit knowledge in natural language, different from the target programming language, which could be indirect. Furthermore, we observed that the presence of docstrings alongside APIs does not consistently enhance performance when compared to using APIs alone, despite occasional cases where such combined prompts achieve the highest scores. This discrepancy may be attributed to the long text of the docstrings, which sometimes hinders the comprehension of APIs when incorporated into the prompt.

%Comparing all knowledge-enhanced prompts, using API sequence and docstring as prompt reaches the highest performance gain, suggesting that these two types of knowledge do have a complementary effect when combined\gu{perhaps we can rewrite this to emphasize that although it reaches the highest performance, the improvement is insignificant (In response to Review 1.5 and 1.6)}.

For two combined prompts that have the same knowledge elements but in a different order, we can see they have similar results for code generation enhancement, but API sequence + docstring prompt is slightly better, indicating that the orders in which knowledge is organized do affect the results.

\begin{tcolorbox}[width=\linewidth, boxrule=0pt, sharp corners=all,
 left=2pt, right=2pt, top=2pt, bottom=2pt, colback=gray!20]
\textbf{Finding 2}: The performance of domain-specific code generation can be improved by prompting LLMs with domain knowledge, particularly the usage of domain libraries.
\end{tcolorbox}

\section{Code Generation with Domain Knowledge Integration}
%\hy{maybe the title itself is not a RQ, under which we may have some RQs}
In the previous RQs, we have demonstrated that injecting API knowledge into prompts has a positive effect on code generation. However, in real-world application scenarios, the model needs to perform the prediction based on the input code context alone, which does not contain any prior API knowledge. 
This sparks an idea of automatically incorporating knowledge into code generation: we can extend code LLMs by constantly inquiring about API knowledge from a knowledge base and using it to prompt the next code fragments. 
In this section, we experiment with three strategies to integrate knowledge into code-generation LLMs and compare their effectiveness. Specifically, we aim to address the following research questions:

\textbf{RQ3}: How to effectively integrate domain knowledge in the code generation process? 

\textbf{- RQ3.1}: Is API recommendation useful for automatically prompting LLMs to generate domain-specific code?

\textbf{- RQ3.2}: Is chain-of-thought useful for automatically prompting LLMs to generate domain-specific code?

\textbf{- RQ3.3}: By fine-tuning, can we further enhance the effectiveness of chain-of-thought for LLMs?

We introduce the three experimental strategies in the following sections and then present our empirical results.

\subsection{Prompting LLM by Inquiring External Knowledge} 

\begin{figure*}[htbp]
    \centering
    \subcaptionbox{Inquiring external knowledge GPT (kg-GPT)}{
        \includegraphics[width=0.8\columnwidth, trim=0 10 0 20 clip]{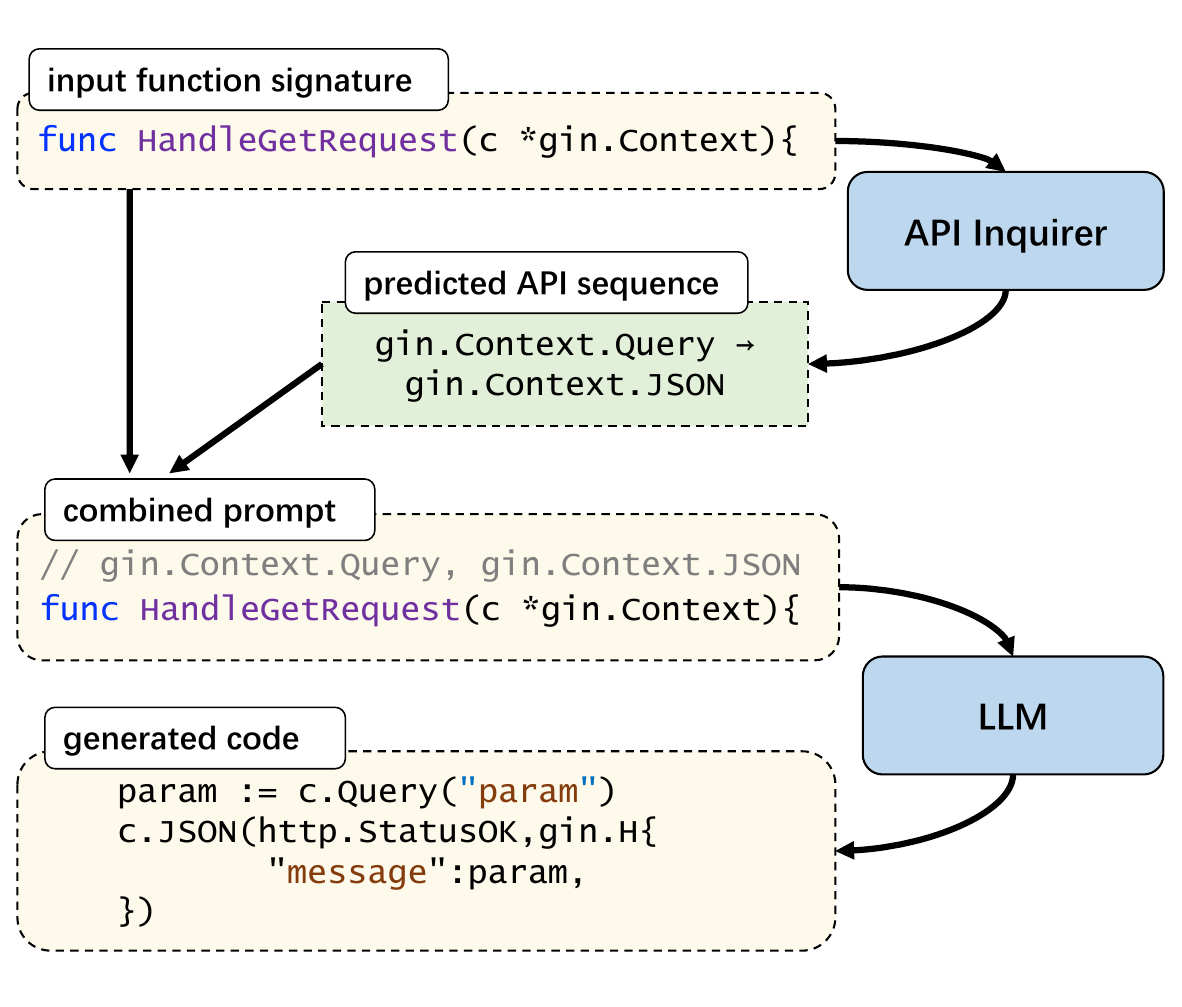}
    }
    \subcaptionbox{Chain-of-thought prompting (CoT-PT)}{
        \includegraphics[width=0.9\columnwidth, trim=-50 0 0 0 clip]{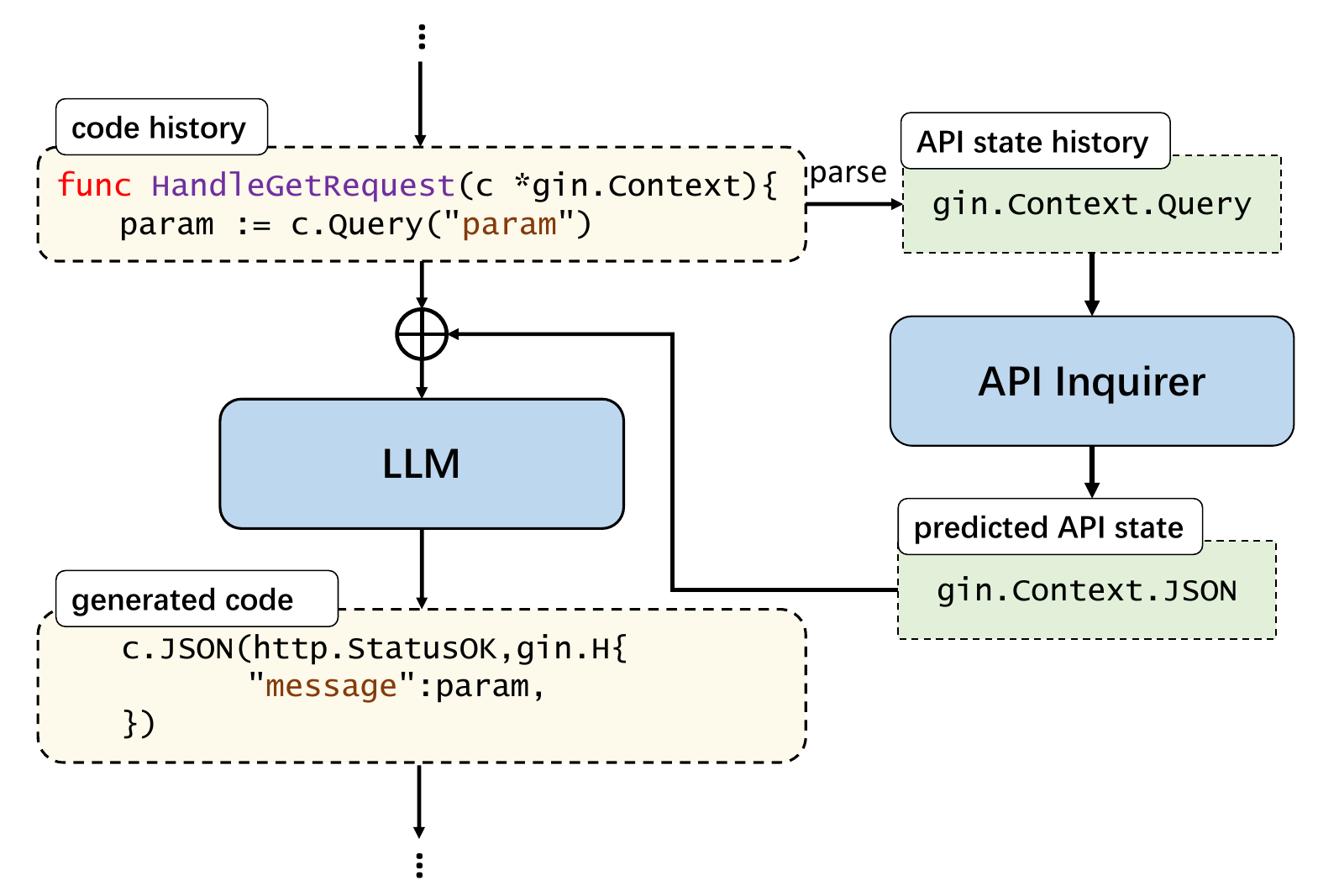}
    }
    \subcaptionbox{Chain-of-thought fine-tuning (CoT-FT)}{
        \includegraphics[width=0.5\textwidth, trim=0 0 0 -10 clip]{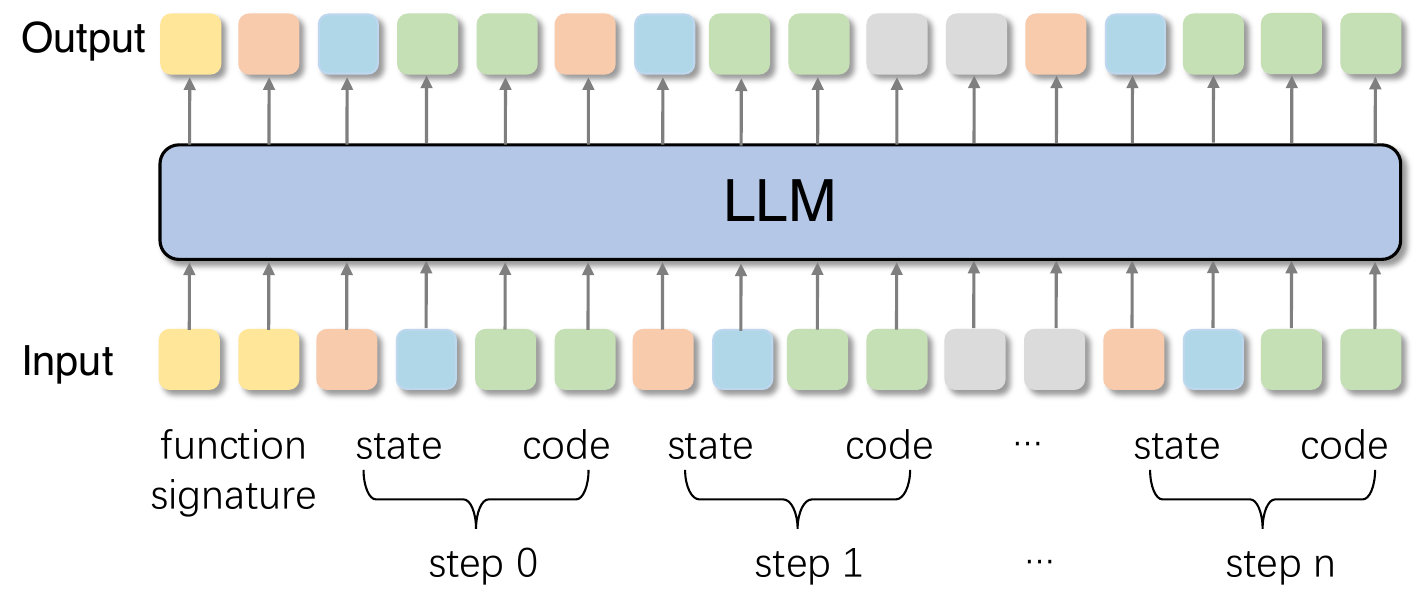}
    }
    \caption{Illustration of the three experimented strategies for knowledge integration}
    \label{fig:strategies}
\end{figure*}

Perhaps the most straightforward approach for injecting knowledge into a code LLM is to train an external knowledge inquiring model named kg-GPT. 
This strategy consists of two modules, one for API knowledge inquiry and the other for code generation, as illustrated in Figure~\ref{fig:strategies} (a). Given the input function signature, the API knowledge inquirer recommends a sequence of possible APIs that could be used to implement a function. Then, the LLM takes the predicted API sequence and the input function signature as a prompt to generate the function body.

The API inquirer is realized as a GPT trained using causal language modeling, each training sequence is composed of a function signature and its corresponding API usages. In this way, it can generate a list of APIs by predicting the next sequence when given the function signature as a context.
%The code generation module is a GPT model which has been pretrained on a code corpus to perform general-purpose code generation.
To introduce the domain-specific API knowledge, we add the predicted API list as a line of comment above the function signature, forming a combined prompt. We then feed the hint to the LLM, which generates the function body.
Owing to the API knowledge in the prompts, it inclines to generate code containing the recommended APIs.

\subsection{Prompting LLMs with Chain-of-thought}

While kg-GPT accomplishes the initial goal of incorporating domain knowledge into LLMs, it does so in a shallow fashion. We hypothesize that a more fine-grained integration of domain knowledge into the generation process has the potential to improve the overall generation performance. Inspired by the chain-of-thought prompting~\cite{Chain-of-Thought} for LLMs, we experiment with the second strategy that incorporates knowledge into the code generation process in a chain-of-thought (CoT) manner.

%Chain-of-Thought Prompting has proved to be useful for improving the reasoning capabilities of large language models\hy{cite}. 

\subsubsection{Programming as a Chain-of-thought Process}
%To deal with a complex programming task such as implementing a function, one often adopts a multi-step thinking strategy, which is to break down the whole task into a series of sub-tasks and implement them step by step \cite{}. At each step, one decides what action should be taken based on the history of former steps, then chooses the API that should be used based on one's knowledge of the domain-specific libraries. This thinking process is known as \textit{chain-of-thought.} Chain-of-Thought prompting has proved to be useful for improving the reasoning capabilities of large language models \cite{Chain-of-Thought}.

To deal with a complex task, one often adopts a multi-step thinking strategy, which is to break down the original task into a series of sub-tasks and solve them step by step. This thinking process is known as \textit{chain-of-thought} \cite{Chain-of-Thought}. Chain-of-thought prompting has been proven to be effective in improving the reasoning capabilities of LLMs~\cite{shi2022language,kojima2022large}.

Like reasoning, programming is also a task that requires careful logical thinking~\cite{yusuf2022accurate}. 
When faced with a programming task to implement a complex requirement, one often thinks in a chain-of-thought manner~\cite{cheng2023prompt}. 
This is natural since a complete code segment (e.g., a function) can usually be decomposed into multiple sub-activities \cite{SunXCBW019}, each containing one or a few statements. 
We refer to a sub-activity as a \textit{step} in the coding process. At each step, one decides what action should be taken based on former contexts and how it should be implemented. 
%In domain-specific programming, the implementation is largely a matter of choosing an API. The choice of an API is based on the knowledge of domain-specific libraries, including API names and descriptions, which we refer to as domain-specific knowledge.
For domain-specific code, this decomposition of activities is much related to API calls of third-party libraries~\cite{apirec,NguyenRRP20}. An API acts like an intermediary, providing developers with functionality implemented by a third-party library. One statement calling an API can be regarded as a separable sub-activity.

%Overall, the thinking process for domain-specific coding is a chain-of-thought procedure integrated with knowledge.

\begin{figure}[htbp]
    \centering
    \includegraphics[width=0.8\columnwidth, trim=0 0 0 0 clip]{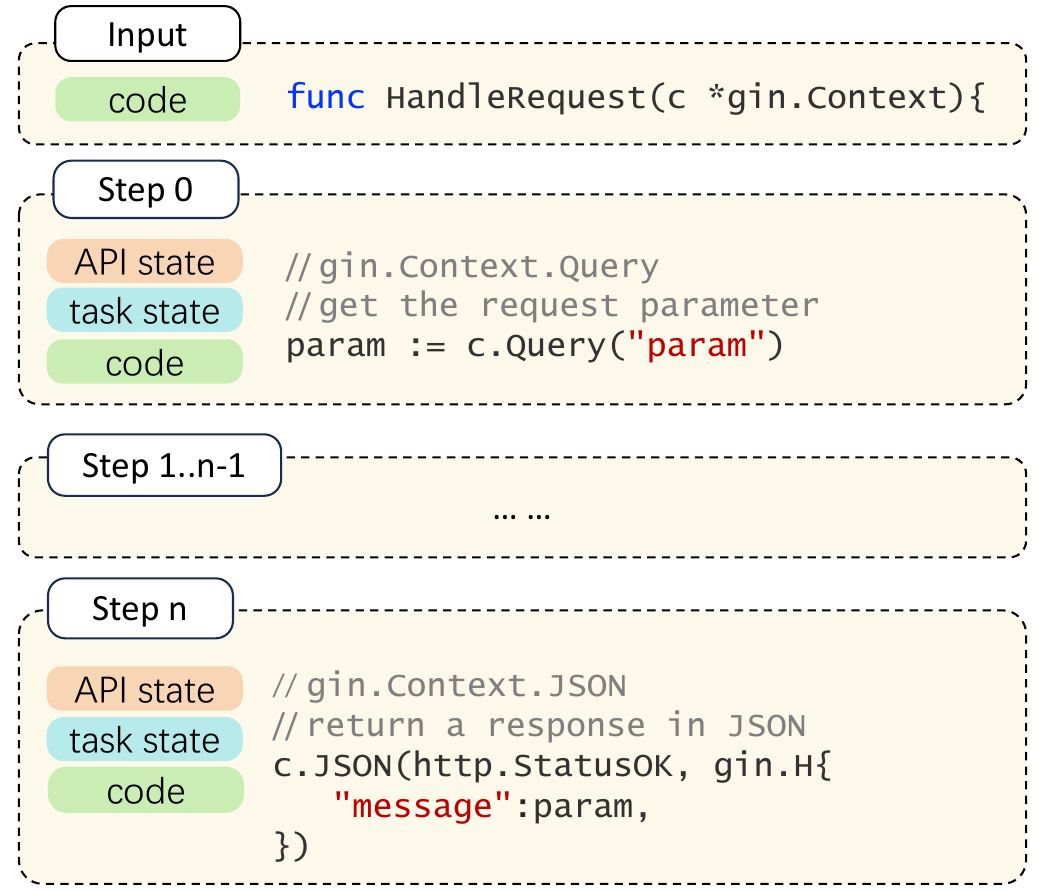}
    \caption{A chain-of-thought view of code generation}
    \label{CoT_example}
\end{figure}

%In this strategy, we aim to simulate this chain-of-thought process in code generation, leading the model to generate a few knowledge statements to perform a sub-task at each step and repeat this action until the whole programming task is finished. The model also additionally generates 'knowledge states' at each step, which act as assistance to the thought process.
Having formulated code generation as a chain of thinking steps, we attempt to integrate knowledge into each step through prompting.
We introduce \textit{knowledge states} to simulate the implicit states in the thinking process.
A knowledge state corresponds to a short code sub-segment, it consists of an API state and a task state. The API state suggests one or more APIs to be used to accomplish this sub-task, and the task states describe the action to be performed. Combined together, they summarize the chain of operations bound to each step.

%In the generation process, a knowledge state is predicted on the basis of the history of all former steps, and it serves as a hint to the code generation of the current step, and also a supplement to domain-specific knowledge.
The generation process involves three subtasks: understanding current input, including formerly written code and corresponding states, predicting the next knowledge state, and generating the next code snippet. The three steps are executed in a loop until we have generated a complete function body or the maximum generation length is reached.

Figure \ref{CoT_example} shows an example of a function generation process in a chain-of-thought manner. 
We take a function signature as input, where the function name \texttt{\small HandleGetRequest} indicates that the requirement is to handle GET requests, and the parameter \texttt{\small c *gin.context} points to the \textit{Gin} library. Based on the input, the model generates code step by step. 
At step 0, it first predicts the API state \texttt{\small gin.Context.Query} and the task state \textit{get the request parameter}. The states suggest that the following code should perform the task of retrieving the request parameter using the API \texttt{\small gin.Context.Query}. Then conditioned on the code history (which is the function signature for step 0) and the current state, the model generates code for step 0, \texttt{\small param := c.Query(``param")}.
Similarly, the model continues to recursively predict code with the states at each step, until the end of the function is reached. 
%\hy{what if no API is involved in a state?}
If no API is required, we mark it as an empty state and simply skip the API prediction for this step.

% Chain of Thought Prompting can improve the reasoning capabilities of large language models. Therefore we introduce the chain-of-thought methodology into the code generation process.

% Specifically, we treat code generation as a multi-step problem and break it down into intermediate steps. For each step, we keep a knowledge state.

% \noindent\textbf{Causal Language Modeling}

%Having decomposed the code generation process, we have two ways to integrate the domain knowledge in the chain-of-thought fashion: in one way, we follow the previous works to directly prompt LLMs in each generation step, which we denoted as \textit{chain-of-thought prompting}. We can also integrate the chain-of-thoughts more deeply through fine-tuning the LLMs, which we denoted as \textit{chain-of-thought finetuning}. Figure \ref{training} illustrates the two ways.
 
%\noindent\textbf{Chain-of-Thought Prompting:}
\subsubsection{Overall Design}

%Following prior work\cite{Chain-of-Thought}, we randomly sampled a set of functions from the corpus as exemplars, processed them by adding knowledge states, and use the fixed set of processed exemplars\gu{this does not seem to be related to the abovementioned API states.} for each code generation problem.

%Since we have decomposed the code generation process into multiple steps, we employ fine-grained prompting by using API states as prompts to guide each step.
Figure~\ref{fig:strategies} (b) illustrates the design of this strategy. It consists of two modules, one for prompt generation and the other for code generation. 
%We define code history as the input function signature and all code generated in previous steps, and API state history as all the APIs used in code history.
The prompt generator is realized with an external knowledge enquiring model kg-GPT as shown in Figure \ref{fig:strategies} (a), it predicts the next possible API based on the API state history or input function signature. 
For each step, the prompt-generator predicts an API state based on the state history. The corresponding API docstring is appended as the task state. The generated API and task states are combined with the code history and used as a prompt for the LLM, which then predicts code for the current step.

\subsection{Chain-of-Thought Fine-tuning}

When prompting with chain-of-thought, the prompt-generation module relies only on API state history for prediction, which lacks information on code syntax and structure, affecting the prediction accuracy. To improve the quality of predicted APIs, we experiment with the third strategy to predict API state conditioning on code history, namely, chain-of-thought fine-tuning (CoT-FT), as illustrated in Figure~\ref{fig:strategies} (c).  In this strategy, we combine all sub-steps together and finetune a single language model in an end-to-end fashion. For each function, we construct a training sequence, which consists of a user input function signature and an output code segment decomposed as $n$ steps. 
At a step \(t\), the model reads the history \(H_{t-1}\), then generates a knowledge state \(K_t\) and a code snippet \(C_t\), concatenated as $S_t = [ K_t, C_t]$.
The history includes the user input \(I_0\) and all previously generated $t$$-$1 steps \([S_0, S_1, ... S_{t-1}]\).
\[H_{t-1} = [I_0, S_0, S_1, ... S_{t-1}]\]
Conditioned on the history, the model first predicts a knowledge state \(K_t\), which consists of an API sequence and a task description.
\[K_t = LM(H_{t-1})\]
This knowledge state is then appended to the history, forming a complete context. The model conditions on this input to generate code for step \(t\).
\[C_t = LM([H_{t-1},K_t])\]
Combining each step, the overall code generation process is shaped as: 
\[K_0,C_0,K_1,C_1,...K_n,C_n = LM(I_0)\]
After removing the states, the final generated code is obtained:
\[ C = C_0,C_1...C_n\]

We adopt the end-to-end training method by combining all steps as a single training sequence and fine-tuning a causal language model (i.e., Transformer decoder)~\cite{VaswaniSPUJGKP17}. A complete training sequence is concatenated by one user input and all sub-sequences, shaped like 
$[ I_0, \ S_0, \ S_1 \ ... \ S_n ]$.
By causal language modeling, the model is trained to predict the probability of the next token based on all former tokens. In this way, we can model the joint probability over the entire sequence.

\subsection{Experimental Setup}
\label{rq3setup}
\noindent\textbf{Collecting API Knowledge}
To evaluate code generation with domain knowledge, we extend the domain-specific dataset for RQ1 and RQ2 (Section~\ref{sec:data}) with domain knowledge. We \rev{are concerned with} knowledge about certain libraries, including the APIs and their corresponding natural language docstring, organized in a dictionary format.
As Table~\ref{table:api_knowledge} depicts, an API name is a unique identifier of a specific function in a third-party library, and \rev{an API docstring} provides a supplementary and more detailed explanation of its behavior.

 \begin{table}[htbp]  
   % \small
    \centering
    \caption{Examples of API knowledge}
    \scalebox{0.9}{
        \begin{tabular}{ll}  
            \toprule
            \bf API name &\bf  API docstring \\
            \midrule
            % \hline
             gin.BasicAuth  &  Returns a Basic HTTP Authorization middleware \\
             gin.Context.Abort & Prevents pending handlers from being called \\
             gin.Context.JSON & Serializes a given struct as JSON \\%into the response body \\
            \bottomrule
        \end{tabular}
    }
    % }
    \label{table:api_knowledge}
\end{table}  

 We obtain API calls from the collected functions in the original dataset. For each function, we parse the code using \textit{tree-sitter}\footnote{https://tree-sitter.github.io/tree-sitter/}. We identify all <variable name, variable type> pairs in a function from the parameter definition and variable declaration statements, if the variable type is a struct or a class defined in the target library, we mark the corresponding variable as targeted.
 We then identify all call expression statements and extract the names of the called function.
 If the call expression contains our target package/class identifier or the called function is initiated by a target variable, we mark it as a target API. API names are then used as queries to retrieve document strings from the knowledge base.

 We obtain API docstring either from the API documentation or from the library source code.
 For libraries with comprehensive official API reference documentation, including \texttt{Unreal Engine}, \texttt{cocos2d-x}, and \texttt{bgfx}, we extract knowledge directly from the document by crawling the web page, then keep the pairwise API declarations and descriptions.
 For libraries without a document, including \texttt{Gin}, \texttt{gRPC-go}, and \texttt{Prometheus}, we use \textit{tree-sitter} to parse the source code of the library, extracting each function with a corresponding doctring if there exists one. We use the function name as the API name, and docstring as the API description.

The parsed API calls and docstrings are integrated into each function for the chain-of-thought fine-tuning.
To implement knowledge integration while conforming to the programming language format, we add the API call and docstrings as comments above each corresponding call expression statement.

 \begin{table*}[tbp]  
    \centering
    \caption{Performance of various API knowledge integration strategies (CB denotes CodeBLEU)%\hy{are these results good or bad. actually, in finetuning, the results of PolyCoder and DomCoderCOT are similar- do the results demonstrate the superiority of the proposed approach? Also, the absolute BLEU values are very low - less than 10 in many cases. I think there are different ways/packages to produce BELU values. BLEU values less than 1 are like random generation? https://cloud.google.com/translate/automl/docs/evaluate}}\gu{Can we make it a pure empirical study rather than claim a novel approach?}%\hy{I think you (and the student) can double check your code for computing BLEU and CodeBLEU. Maybe the code is wrong. There are many strange results in this table. I remember there are different Python packages to compute BLEU in different ways}
    %\hy{are these results good or bad. Also, On some datasets, the results of BLEU and CodeBLEU diff a lot - as high as 10x, which may be unusual...need double check the results.}
    }
        \scalebox{1.0}{
        \begin{tabular}{lr@{}c@{}c@{}c@{}c@{}c@{}c@{}c@{}c@{}c@{}c@{}c@{}}  
            \toprule
            \multirow{2}{*}{\bf Model} & \multicolumn{2}{c}{~~~~~gin~~~~~} & \multicolumn{2}{c}{prometh.} & \multicolumn{2}{c}{grpc-go} & \multicolumn{2}{c}{UnrealEng.}  & \multicolumn{2}{c}{cocos2d-x} & \multicolumn{2}{c}{bgfx}\\
                &\bf BLEU & \bf \, CB & \bf BLEU\, & \bf CB  & \bf BLEU\, & \bf CB & \bf BLEU  &\bf CB & \bf BLEU\, & \bf CB  & \bf BLEU\, & \bf CB\\
            \midrule
             % \ PolyCoder (160M) & 8.13 & 17.05 &  1.77 & 12.11 & 5.37 & 14.84 & 0.76 & 7.16  & 2.00 & 10.24 & 0.94 & 9.89\\
             \ PolyCoder (zero-shot)  & 8.13\,  &  \,17.05 & 1.77 &  12.11 & 55.36 & 57.52 & 0.94 & 9.87 & 13.83  & 23.83 & 0.76 & 7.16 \\
             \ +In-context learning~\cite{ahmed2022fewshotcodesum} & 8.18\, & 16.36 & 2.01  & 12.57  & 55.54 & 57.11  & 0.96  & 0.98 & 13.27 & 23.00 &  0.83 & 7.24  \\
            % \ \rev{+CodeGen4Libs} &  &  &   &   &   &   &  &  &   &  &   &  \\
             \ +\ourmethod (kg-GPT) &\bf 9.91\, & 18.16 & \bf 2.29 & 12.76 & \bf 56.87 &\bf 58.30 &  1.26 & \bf 10.20 & \bf 15.33 &\bf  24.09 & \bf 0.78 & 7.34 \\
             \ +\ourmethod (CoT-PT) & 9.17\, & 17.41 &  2.21 &\bf 12.65 &  55.97  & 57.70 &  \bf1.38 &  10.15 &  16.13 & 24.91 & \bf 0.91 & \bf 7.87\\ \hdashline
             \ PolyCoder (fine-tuning)   & 20.62\, & 27.18 & 5.88 & 17.11 & 62.97 & 63.37 &  1.57 & 11.10 & 27.65 & 32.67 & 1.01 & 11.80 \\
            \ +\ourmethod (CoT-FT)  & \bf21.12\,  &  \bf27.51 & \bf7.59 & \bf19.13 & \bf63.23 & \bf63.83 & \bf 2.26 & \bf11.69 & \bf29.30 & \bf33.99 & \bf1.05 & \bf 11.95\\
            % \ourmethod (API + doc) & \bf21.35  &   \bf28.22  & 6.36 & 18.27 & \bf63.38 & \bf63.95 &  1.78 & 11.43 & 28.67 & 32.40 & 0.91 & 11.51 \\
            \hline
             \ StarCoder (zero-shot) & 8.46\, &17.55 &1.68 &12.37 &58.69 &60.09 &0.58 &9.00 &14.08 &23.56 &0.80 &7.21 \\
            \ +In-context learning~\cite{ahmed2022fewshotcodesum} & 7.11 & 17.32 & 2.00 & 12.75  &  58.12 & 59.07  & 1.14 & 10.76 & 18.62  & 26.55 & 1.12  & 7.93 \\
             %\ \rev{+CodeGen4Libs} &  &  &   &   &   &   &  &  &   &  &   &  \\
             \ +\ourmethod (kg-GPT) &\bf 13.12\, &\bf 20.44 &\bf 3.75 & 14.74 &\bf 60.41 &\bf 61.85 &\bf 1.74 &\bf 10.98 &\bf 22.43 &\bf 27.53 & \bf1.20 & \bf8.67 \\
             \ +\ourmethod (CoT-PT) & 12.46\, & 19.77 & 3.32 & 13.66 & 59.91 & 61.25 & 1.64 & 10.96 & 20.78 & 25.81 & 1.17 & 8.42 \\
             \hline
             \ CodeLlama (zero-shot)
             & 8.47\, & 18.07 & 1.79 & 13.13 &  57.61 & 60.39 & 0.73 & 10.21 & 16.76 & 25.87 & 0.55 & 7.72 \\
             \ +In-context learning~\cite{ahmed2022fewshotcodesum} & 8.22\, & 18.58 &  2.69 & 13.87  &  \bf 59.35 & \bf 61.49  & 0.81 & 10.35 & 17.90  & 27.06 & 0.60  & 8.47 \\
             %\ \rev{+CodeGen4Libs} &  &  &   &   &   &   &  &  &   &  &   &  \\
             \ +\ourmethod (kg-GPT) & 11.71\, & 19.83 & 2.62 & 13.41 & 55.93 & 58.24 & 0.99 & 10.55 & \bf 19.58 & \bf 27.71 & \bf 0.65 & \bf 8.66 \\
             \ +\ourmethod (CoT-PT) & \bf 11.75\, & \bf 19.90 & \bf 3.07 & \bf 14.58 & 56.03 & 58.69 & \bf 1.15 & \bf 11.15 & 16.39 & 25.16 & 0.57 & 8.21 \\
            \bottomrule
        \end{tabular}
        }
        % }
    \label{table_rq3}
\end{table*}

 \begin{table*}[tbp]  
    \centering
    \caption{Performance of kg-GPT in API recommendation}
        \scalebox{0.9}{
        \begin{tabular}{lc@{}cc@{}cc@{}cc@{}cc@{}cc@{}c}  
            \toprule
            \multirow{2}{*}{\bf \ Model} & \multicolumn{2}{c}{gin} & \multicolumn{2}{c}{prometheus} & \multicolumn{2}{c}{grpc-go} & \multicolumn{2}{c}{Unreal Engine}  & \multicolumn{2}{c}{cocos2d-x} & \multicolumn{2}{c}{bgfx}\\
                &\bf BLEU\,  &\bf HitRatio & \bf BLEU\, & \bf HitRatio  & \bf BLEU\, & \bf HitRatio &\bf BLEU\,  &\bf HitRatio & \bf BLEU\, & \bf HitRatio  & \bf BLEU\, & \bf HitRatio\\
            \midrule
             \ PolyCoder  & 16.81     & 0.03  & 8.02   & 0.03 & 32.09  & 0.37  & 7.47     & 0.04     & 30.10     & 0.33      & 3.32     & 0.01  \\
             %\ kg-GPT &\bf 45.33 & \bf0.36 & \bf23.99 & \bf0.20 & \bf52.64 & \bf0.67 & \bf15.09  & \bf0.09 & \bf56.02 & \bf0.50 & \bf31.02 & \bf0.27\\
            %\hdashline
            \ CodeLlama  & 21.64    & 0.06  & 12.42   & 0.04 & 36.92  & 0.38  & 9.50  & 0.07 & 35.91  & 0.36      & 6.73 & 0.03 \\
             \ kg-GPT &\bf 46.48 & \bf0.36 & \bf27.16 & \bf0.20 & \bf52.67 & \bf0.67 & \bf16.48  & \bf0.09 
 & \bf57.24 & \bf0.50 & \bf34.25 & \bf0.28\\
            \bottomrule
        \end{tabular}
        }
        % }
    \label{table:result:kg-gpt}
\end{table*}

\noindent\textbf{Backbone Models}
 We employ PolyCoder \cite{xu2022polycoder}, StarCoder~\cite{li2023starcoder}, and CodeLlama \cite{roziere2023codellama} as backbone LLMs, which are well-established open-source models trained on GitHub code. Considering our computational resources, all fine-tuning experiments were only carried out on PolyCoder. PolyCoder is based on GPT-2, which is easy to develop and deploy. More importantly, it demonstrates competitive performance to GPT-3 counterparts~\cite{xu2022polycoder}.  
We employ the released checkpoint of pre-trained PolyCoder-2.7B, StarCoder-15.5B\footnote{https://huggingface.co/bigcode/starcoder}, and CodeLlama-7B\footnote{https://huggingface.co/codellama/CodeLlama-7b-hf} for all experiments. 
We implement the API inquirer (kg-GPT) by employing a standard GPT-2 model with $12$ layers and $768$ dimensionality. 

\noindent\textbf{Training and Prediction}
We train all models in a server with 8 Nvidia A100 (40GB) GPU cards. To fine-tune the backbone PolyCoder, we follow the hyper-parameter configurations in the original paper~\cite{xu2022polycoder}, with a batch size of $32$ and a learning rate of 1.6$e$-4. We train the API inquirer with a batch size of $16$ and a learning rate of 5$e$-5.

We randomly sample 500 functions %\gu{explain the reason for 500 in response to Review 1.10 (a)} 
from each dataset to conduct the prediction and evaluation. For each function, we configure the model to generate new tokens until we either encounter the termination token <EOF> or reach a maximum token length of 256, then truncate the output to the end of the function. 
For the chain-of-thought models, we do not count the tokens occupied by knowledge states in the overall length. 

\noindent\textbf{Comparison Methods} 
We compare our approach with all backbone LLMs in the zero-shot setting, which refers to prompting the pre-trained checkpoint without any further fine-tuning. %We also compare \ourmethod with CodeGen4Libs~\cite{2023liucodegen4libs}, a related work on code generation for specific libraries. 
We also compare \ourmethod with a few-shot learning method proposed by Ahmed et al.~\cite{ahmed2022fewshotcodesum}, which prepend a few in-context examples in the LLM context. We randomly sample 10 in-context examples that invoke the same library to the target domain.  
Besides the zero-shot setting, we are also concerned with the fine-tuning strategy of our method. Specifically, we compare \ourmethod (CoT-FT) with PolyCoder fine-tuned on the domain-specific dataset.
%%We also experiment with two variations of the chain-of-thought approach: API fine-tuning and API+docstring finetuning. 
%%We take two baselines: \textit{zero-shot PolyCoder}, which refers to using plain function signatures as prompts.
%%and \textit{PolyCoder-ICL}, which applies In-Context Learning to the model. 

\begin{figure}[htp]
    \centering
    \includegraphics[width=1\columnwidth, trim=20 68 20 40 clip]{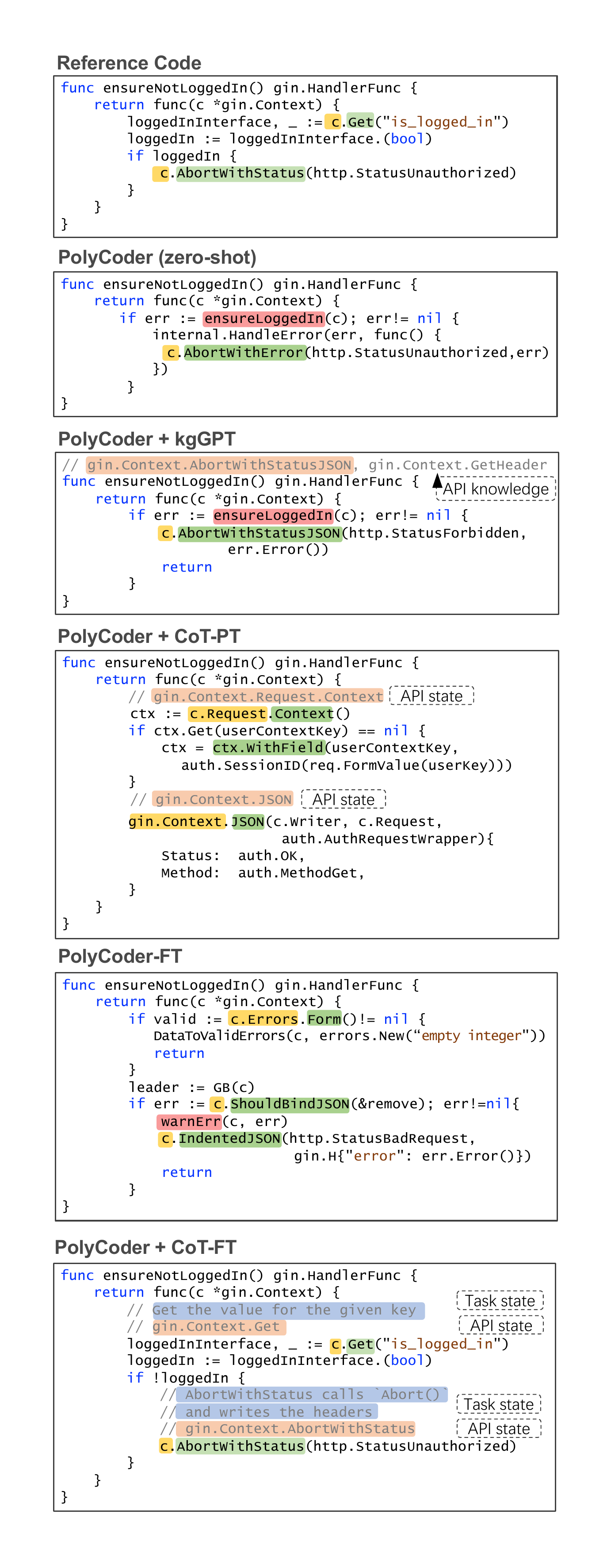}
    \caption{Examples of generated code%\cm{we displayed finetune+API results only, inconsistent with the API+doc in figure}
    }
    \label{fig:case1}
\end{figure}

\subsection{Results}

The evaluation results for all methods and strategies are shown in Table~\ref{table_rq3}. 
Under both the zero-shot and the fine-tuning settings, BLEU and CodeBLEU scores rise when the model is knowledge-enhanced, and the chain-of-thought fine-tuning strategy reaches the best scores.
%\hy{in Table 7. for the PolyCoder, can briefly say what it is (original model, does not incorporate the knowledge of API, the same one as used in section 3, etc) }

Under the zero-shot setting, both the kg-GPT and CoT-PT strategies improve the results, with kg-GPT demonstrating better performance. API knowledge introduced by kg-GPT raised the BLEU scores by 49.33\% on average, and the CodeBLEU scores by 9.82\%. 
Under the fine-tuning setting, chain-of-thought raised the BLEU scores by 17.10\% on average, and the CodeBLEU scores by 4.20\%. %The improvement is challenging but substantial, considering that the baseline is a large model with 2.7B parameters. %in a fiercely competitive domain of large language models.
Furthermore, we can see that the performance is improved more significantly on datasets with poor zero-shot results, such as \texttt{prometheus} and \texttt{Unreal Engine}.
Comparatively, the CoT-FT strategy provides modest improvement, except for the CodeBLEU scores for some libraries such as \texttt{gin} and \texttt{prometheus}.
Both kg-GPT and CoT-PT incorporate API knowledge into the zero-shot PolyCoder, the differences in results could be attributed to their distinct knowledge-integration strategies. 
Kg-GPT predicts all states of APIs at once, whereas CoT-PT predicts APIs incrementally based on historical states, which do not contain information about code syntax and structure, resulting in unstable quality of the predicted APIs.

Domain-finetuning improves the results compared to zero-shot, as it allows the model to learn about domain-specific code most straightforwardly. Still, by injecting the knowledge into training data in a chain-of-thought fashion, the performance is raised to a higher level. 
We note that the improvement that CoT-FT brings to PolyCoder is relatively marginal. Nevertheless, it demonstrates that the proposed chain-of-thought knowledge integration is also useful in the fine-tuning mode despite still having ample room for further improvement.

Note that, in most cases, BLEU scores are smaller than CodeBLEU scores. Particularly, in Unreal Engine and bgfx, CodeBLEU score is ten times of BLEU score.
It is caused by the metric design: BLEU only evaluates the n-gram match, while CodeBLEU examines a combination of the n-gram match, syntax match, and dataflow match. In practice, a low n-gram match could come with a high syntax match and dataflow match, especially when the generated code has correct functionality but different variable names. 

As an essential part, we assess the accuracy of the APIs recommended by kg-GPT. This evaluation holds significance since it influences the language model's performance in code generation relying on the suggested API sequences. We employ the same dataset outlined in Section~\ref{rq3setup} for training the kg-GPT. We measure the quality of API recommendations using BLEU and Hit Ratio. 
The Hit Ratio refers to the percentage of correctly recommended APIs out of the total number of predicted APIs. A higher Hit Ratio indicates a better performance of the API recommendation by the kg-GPT. For comparison, we provide the results of API recommendation by PolyCoder-2.7B. These results are obtained by extracting API sequences from the predicted code by PolyCoder. The results are presented in Table~\ref{table:result:kg-gpt}. We obverse that kg-GPT recommends domain API sequences with high accuracy. For example, it achieves 10 times the accuracy in the \texttt{bgfx} library in terms of BLEU score. 

% To illustrate, we present an example in Figure~\ref{codebleu_example}.
% For a function using the Prometheus library,  we compare the zero-shot PolyCoder-generated code with the reference code. We compute the score only for function bodies as the function signature is the given input.
% The n-gram match score is only 0.31\%, as we can see the matched n-grams make up only a small portion of the reference code. 
% However for syntax match, the reference function has 82 subtrees, and 10 of them show up also in the generated code, resulting in a match score of 12.20\%.  
% And the data-flow match score is 50.00\%, as the reference code has 8 DFG edges, and 4 of them match with the generated code.

% \begin{figure}[htbp]
%     \centering
%     \fbox{
%         \includegraphics[width=7cm, trim=0 10 0 0 clip]{figures/codebleu_example.pdf}
%     }
%     \caption{Example of BLEU and CodeBLEU computation\hy{this example shows that the generated code is bad?}\gu{can we show a clearly good example?}}
%     \label{codebleu_example}
% \end{figure} 

%\hy{looking at the sample results of these two metrics, the differences shouldn't be very high? There is also an Abation Study at the bottom: https://github.com/microsoft/CodeXGLUE/blob/main/Code-Code/code-to-code-trans/CodeBLEU.MD}

\subsection{Qualitative Analysis}
\label{sec:case}

We inspected the code generated by various methods and presented cases to illustrate their effectiveness. Figure \ref{fig:case1} presents an example of code generation by various methods.
% \cm{we have presented just one example in figure}

We take a function using the Gin library as an example. As its name indicates, the reference function ensures the logged-in status is False. It checks the logged-in status using \texttt{gin.Context.Get}, and aborts using \texttt{gin.AbortWithStatus} if it's True.

For this function, we compare the results of all strategies.
Among them, the CoT-FT strategy generates the best result. Through step-by-step generation, it first predicts the task description and API name as the knowledge state, then generates code to implement the task in the description using the API. Since it correctly predicts the state at each step, it eventually generates the correct code segment that meets the requirements. 
Comparatively, all other methods give more or less erroneous results. For example, the code generated by the PolyCoder-FT called several APIs of the targeted gin library, but none of them is correct, failing to fulfill the requirements.

%The second case is a function that creates a middleware to record the latency of HTTP requests for a given route using Prometheus. 
%When a request comes in, it receives the current route and path using \texttt{mux.CurrentRoute} and  \texttt{route.GetPathTemplate}, starts a timer using \texttt{prometheus.NewTimer}, calls the next handler using \texttt{next.http}, and then records the duration of the timer using \texttt{timer.ObserveDuration}.
%The code generated by the vanilla PolyCoder simply called \textit{prometheus.NewSummaryMetric}, which does not actually implement the latency recording requirement.  
%The code generated by the CoT-PT strategy\gu{?} contains all correct APIs in the right order.

\begin{tcolorbox}[width=\linewidth, boxrule=0pt, sharp corners=all,
 left=2pt, right=2pt, top=2pt, bottom=2pt, colback=gray!20]
 \textbf{Finding 3}: The performance of domain-specific code generation can be improved through domain knowledge integration in the generation process of LLMs.
\end{tcolorbox}

\subsection{Industry Study}
To fully assess the effectiveness of \ourmethod, we perform an in-house industry study on Tencent Inc, a world-famous IT company. Our study focuses on the domain of messenger Apps (e.g., WeChat, QQ) which are the predominant products of this company. 
%Data
We start by creating a survey about the domain libraries that are mostly adopted by the developers in the group of messenger products. Results show that an internal library called tRPC-go (https://github.com/trpc/trpc) ranked as the most popular one that is utilized by the group.
We therefore selected functions % around a few thousands of functions
that utilize this library and evaluate the performance of \ourmethod in completing these functions. 
To ensure that the selected functions are representative and distinct, we specifically chose functions containing 3 to 7 APIs. Functions with identical API sequences were clustered into groups. These function groups were then sorted based on the number of functions they contained, in descending order. From each class, we randomly selected one function. Additionally, we manually filtered out functions that developers may find challenging to understand. As a result, we obtained a set of 30 test functions for evaluation.

We built \ourmethod based on CodeLlama-7B and applied it to complete the selected test functions. We test the kg-GPT strategy which demonstrates the best performance in the quantitative study and is easier to deploy.
%Metrics:
We measure the performance using two metrics: 1) \textbf{task-relevance}, which means that the generated content satisfies the intention in the function signature, and 2) \textbf{semantic correctness}, which means that the implemented function body is semantically correct. %However, as I do not know enough about the syntax of the trpc-go library, I do not check whether the call parameters and so on are syntax-correct.
We asked three developers from the relevant team in Tencent to rate the evaluation criteria independently. Conflicts were resolved through discussion until a consensus was reached. 

The results are presented in Table~\ref{table:result:industry}. We find that \ourmethod significantly enhances the quality of generated code. The semantic correctness of the generated code by \ourmethod is triple that of the code generated by the original CodeLlama-7B. The results confirm the effectiveness of \ourmethod in wider application scenarios and metrics in the industry.

 \begin{table}[tbp]  
    \centering
    \caption{Performance of \ourmethod in the industry study}
    \begin{tabular}{lcc}  
        \toprule
         \multirow{2}{*}{\bf Model} & \bf Task  & \bf  Semantic \\
           & \bf  Relevance  & \bf Correctness \\
         \hline
         CodeLlama-7B (zero-shot) &  20/30 & 6/30 \\ 
         - w/ \ourmethod (kg-GPT) & 28/30 & 18/30 \\
         \bottomrule
    \end{tabular}
    \label{table:result:industry}
\end{table}

\section{Discussions}

\subsection{Future Directions}
%\gu{1. More knowledge sources such as Stack Overflow 2. More efficient way to integrate knowledge such as knowledge involved pertaining? (take reference to the knowlege+BERT that you worked before) 3. More advanced knowledge enquirer such as a QA system., etc.}

Based on our findings, we suggest three possible future directions for domain-specific code generation by LLMs. 

First, programming knowledge can extend beyond API documentation. We suggest future research to consider more sources of domain knowledge in the design of prompts and the integration of knowledge.
Regarding syntax, more comprehensive information about third-party libraries could be helpful, including class descriptions, relationships between classes, and other relevant information.
For code semantics, descriptive and explanatory information from sources like Stack Overflow could be useful. 
Moreover, knowledge about a certain domain such as workflows and code templates could also be useful.

Second, results on RQ2 demonstrate that docstrings also play an important role in the design of prompts. But this presents a new challenge of increasing the code length. In addition, the ability to understand and generate natural language descriptions is relatively limited for code-oriented models. We suggest future research to design better mechanisms to integrate knowledge into code, bridging the information gap between programming language and natural language. For example, we can consider pre-training tasks that jointly predict code and knowledge. 

The results of the kg-GPT strategy suggest that a separate knowledge enquirer can be an effective way to incorporate knowledge with code. In the future, researchers can focus on designing more sophisticated knowledge inquirers. For example, they can build a QA or search engine that detects any factual knowledge and provides the prompt during the code generation process. %\hy{anything about COT, better utilization of LLM?}

%In our study, we explore the incorporation of API knowledge into code generation in a chain-of-thought (CoT) manner. 
The results also show that chain-of-thought is an efficient way to incorporate knowledge into LLMs in both fine-tuning and usage stages. In the future, researchers could explore integrating knowledge in the early stage of LLM training. Instead of training a casual language model on plain code, they could augment code with API knowledge and use it in pre-training. They could also decompose source code into sub-modules, each representing a specific sub-activity, and generate the modules in a certain order as a simulation of chain-of-thought.

\subsection{Threats to Validity}
\label{sec:threats}

% The threats to validity lie in two aspects.

% \paragraph{Evaluation Metrics} The well-accepted metric for code generation pass@k refers to the passing rate of generated code on test cases. It is commonly used with the HumanEval dataset, which contains hand-crafted test cases for a small number of simple functions. However, we did not adopt this  evaluation metric, since domain-specific functions are more complex and rely on a fully built executable project, leading to a higher workload to create the test cases and run them.

%\paragraph{Model Size} Due to the restriction of computational resources, we experimented with LLMs that have 2.7B parameters. 

%\ \ \ \ \textit{\textbf{Internal validity.}}
%Our study is based on the assumption that domain-specific code commonly relies on specific libraries. However, it is essential to acknowledge that there may exist domains where code is not dependent on such libraries. Consequently, these library-independent domains might be excluded from the scope of our study.\hy{perhaps need not say that, no evaluation either.}

%\textit{\textbf{External validity.}}
The empirical study in this research investigates three LLMs. In particular, the code-specific LLMs we studied are limited to a parameter size not exceeding 2.7B. Considering the rapid development of LLMs, the LLMs studied in this paper may not fully represent all the latest advancements in large-scale models. Furthermore, while our investigation primarily focuses on code generation tasks in the domains of web and game development, it is essential to recognize that code generation is a ubiquitous task across a wide range of domains. Therefore, the findings obtained from the two particular domains can potentially offer general insights applicable in various contexts. Regarding the evaluation metrics, we have utilized BLEU and CodeBLEU. Although these two metrics are widely used in code intelligence research, they may not fully represent human experience.

\section{Conclusion}

In this paper, we conduct an empirical study on domain-specific code generation by LLMs. Our study finds that LLMs such as ChatGPT exhibit sub-optimal performance in generating domain-specific code, and adding knowledge prompts about domain libraries improves the performance. To further investigate how to incorporate API knowledge into LLMs, we design three experimental strategies to automatically prompt LLMs, including inquiring about an external knowledge GPT, chain-of-thought prompting, and chain-of-though fine-tuning. Our experimental results show the effectiveness of knowledge integration in both zero-shot and fine-tuning settings. 
We hope our study can inspire future work on improving the code generation capabilities of LLMs for specific domains. 

%\section*{Data Availability}
%Our source code and experimental data are available at \url{https://anonymous.4open.science/r/DomainCodeGen-D913}.

\section*{Acknowledgment}
This research is supported by the National Key R\&D Program of China (Grant No. 2023YFB4503802), the National Natural Science Foundation of China (Grant No. 62102244), and the CCF-Tencent Open Research Fund (RAGR20220129).

\bibliography{references}

%%% -*-BibTeX-*-
%%% Do NOT edit. File created by BibTeX with style
%%% ACM-Reference-Format-Journals [18-Jan-2012].

\begin{thebibliography}{56}

%%% ====================================================================
%%% NOTE TO THE USER: you can override these defaults by providing
%%% customized versions of any of these macros before the \bibliography
%%% command.  Each of them MUST provide its own final punctuation,
%%% except for \shownote{}, \showDOI{}, and \showURL{}.  The latter two
%%% do not use final punctuation, in order to avoid confusing it with
%%% the Web address.
%%%
%%% To suppress output of a particular field, define its macro to expand
%%% to an empty string, or better, \unskip, like this:
%%%
%%% \newcommand{\showDOI}[1]{\unskip}   % LaTeX syntax
%%%
%%% \def \showDOI #1{\unskip}           % plain TeX syntax
%%%
%%% ====================================================================

\ifx \showCODEN    \undefined \def \showCODEN     #1{\unskip}     \fi
\ifx \showDOI      \undefined \def \showDOI       #1{#1}\fi
\ifx \showISBNx    \undefined \def \showISBNx     #1{\unskip}     \fi
\ifx \showISBNxiii \undefined \def \showISBNxiii  #1{\unskip}     \fi
\ifx \showISSN     \undefined \def \showISSN      #1{\unskip}     \fi
\ifx \showLCCN     \undefined \def \showLCCN      #1{\unskip}     \fi
\ifx \shownote     \undefined \def \shownote      #1{#1}          \fi
\ifx \showarticletitle \undefined \def \showarticletitle #1{#1}   \fi
\ifx \showURL      \undefined \def \showURL       {\relax}        \fi
% The following commands are used for tagged output and should be
% invisible to TeX
\providecommand\bibfield[2]{#2}
\providecommand\bibinfo[2]{#2}
\providecommand\natexlab[1]{#1}
\providecommand\showeprint[2][]{arXiv:#2}

\bibitem[Ahmed and Devanbu(2022)]%
        {ahmed2022fewshotcodesum}
\bibfield{author}{\bibinfo{person}{Toufique Ahmed} {and}
  \bibinfo{person}{Premkumar Devanbu}.} \bibinfo{year}{2022}\natexlab{}.
\newblock \showarticletitle{Few-shot training llms for project-specific
  code-summarization}. In \bibinfo{booktitle}{\emph{Proceedings of the 37th
  IEEE/ACM International Conference on Automated Software Engineering}}.
  \bibinfo{pages}{1--5}.
\newblock


\bibitem[AI(2023)]%
        {ChatGPT}
\bibfield{author}{\bibinfo{person}{Open AI}.} \bibinfo{year}{2023}\natexlab{}.
\newblock \bibinfo{title}{ChatGPT}.
\newblock \bibinfo{howpublished}{\url{https://openai.com/blog/chatgpt}}.
\newblock


\bibitem[Austin et~al\mbox{.}(2021)]%
        {austin2021mbpp}
\bibfield{author}{\bibinfo{person}{Jacob Austin}, \bibinfo{person}{Augustus
  Odena}, \bibinfo{person}{Maxwell Nye}, \bibinfo{person}{Maarten Bosma},
  \bibinfo{person}{Henryk Michalewski}, \bibinfo{person}{David Dohan},
  \bibinfo{person}{Ellen Jiang}, \bibinfo{person}{Carrie Cai},
  \bibinfo{person}{Michael Terry}, \bibinfo{person}{Quoc Le}, {et~al\mbox{.}}}
  \bibinfo{year}{2021}\natexlab{}.
\newblock \showarticletitle{Program synthesis with large language models}.
\newblock \bibinfo{journal}{\emph{arXiv preprint arXiv:2108.07732}}
  (\bibinfo{year}{2021}).
\newblock


\bibitem[Black et~al\mbox{.}(2022)]%
        {black2022gptneox}
\bibfield{author}{\bibinfo{person}{Sidney Black}, \bibinfo{person}{Stella
  Biderman}, \bibinfo{person}{Eric Hallahan}, \bibinfo{person}{Quentin
  Anthony}, \bibinfo{person}{Leo Gao}, \bibinfo{person}{Laurence Golding},
  \bibinfo{person}{Horace He}, \bibinfo{person}{Connor Leahy},
  \bibinfo{person}{Kyle McDonell}, \bibinfo{person}{Jason Phang},
  {et~al\mbox{.}}} \bibinfo{year}{2022}\natexlab{}.
\newblock \showarticletitle{GPT-NeoX-20B: An Open-Source Autoregressive
  Language Model}. In \bibinfo{booktitle}{\emph{Proceedings of BigScience
  Episode\# 5--Workshop on Challenges \& Perspectives in Creating Large
  Language Models}}. \bibinfo{pages}{95--136}.
\newblock


\bibitem[Black et~al\mbox{.}(2021)]%
        {gpt-neo}
\bibfield{author}{\bibinfo{person}{Sid Black}, \bibinfo{person}{Leo Gao},
  \bibinfo{person}{Phil Wang}, \bibinfo{person}{Connor Leahy}, {and}
  \bibinfo{person}{Stella Biderman}.} \bibinfo{year}{2021}\natexlab{}.
\newblock \bibinfo{booktitle}{\emph{{GPT-Neo: Large Scale Autoregressive
  Language Modeling with Mesh-Tensorflow}}}.
\newblock
\urldef\tempurl%
\url{https://doi.org/10.5281/zenodo.5297715}
\showDOI{\tempurl}
\newblock
\shownote{{If you use this software, please cite it using these metadata.}}.


\bibitem[Brown et~al\mbox{.}(2020)]%
        {brown2020gpt3}
\bibfield{author}{\bibinfo{person}{Tom Brown}, \bibinfo{person}{Benjamin Mann},
  \bibinfo{person}{Nick Ryder}, \bibinfo{person}{Melanie Subbiah},
  \bibinfo{person}{Jared~D Kaplan}, \bibinfo{person}{Prafulla Dhariwal},
  \bibinfo{person}{Arvind Neelakantan}, \bibinfo{person}{Pranav Shyam},
  \bibinfo{person}{Girish Sastry}, \bibinfo{person}{Amanda Askell},
  {et~al\mbox{.}}} \bibinfo{year}{2020}\natexlab{}.
\newblock \showarticletitle{Language models are few-shot learners}.
\newblock \bibinfo{journal}{\emph{Advances in neural information processing
  systems}}  \bibinfo{volume}{33} (\bibinfo{year}{2020}),
  \bibinfo{pages}{1877--1901}.
\newblock


\bibitem[Cao et~al\mbox{.}(2023)]%
        {cao2023study}
\bibfield{author}{\bibinfo{person}{Jialun Cao}, \bibinfo{person}{Meiziniu Li},
  \bibinfo{person}{Ming Wen}, {and} \bibinfo{person}{Shing-chi Cheung}.}
  \bibinfo{year}{2023}\natexlab{}.
\newblock \showarticletitle{A study on prompt design, advantages and
  limitations of chatgpt for deep learning program repair}.
\newblock \bibinfo{journal}{\emph{arXiv preprint arXiv:2304.08191}}
  (\bibinfo{year}{2023}).
\newblock


\bibitem[Chen et~al\mbox{.}(2021)]%
        {chen2021codex}
\bibfield{author}{\bibinfo{person}{Mark Chen}, \bibinfo{person}{Jerry Tworek},
  \bibinfo{person}{Heewoo Jun}, \bibinfo{person}{Qiming Yuan},
  \bibinfo{person}{Henrique~Ponde de Oliveira~Pinto}, \bibinfo{person}{Jared
  Kaplan}, \bibinfo{person}{Harri Edwards}, \bibinfo{person}{Yuri Burda},
  \bibinfo{person}{Nicholas Joseph}, \bibinfo{person}{Greg Brockman},
  \bibinfo{person}{Alex Ray}, \bibinfo{person}{Raul Puri},
  \bibinfo{person}{Gretchen Krueger}, \bibinfo{person}{Michael Petrov},
  \bibinfo{person}{Heidy Khlaaf}, \bibinfo{person}{Girish Sastry},
  \bibinfo{person}{Pamela Mishkin}, \bibinfo{person}{Brooke Chan},
  \bibinfo{person}{Scott Gray}, \bibinfo{person}{Nick Ryder},
  \bibinfo{person}{Mikhail Pavlov}, \bibinfo{person}{Alethea Power},
  \bibinfo{person}{Lukasz Kaiser}, \bibinfo{person}{Mohammad Bavarian},
  \bibinfo{person}{Clemens Winter}, \bibinfo{person}{Philippe Tillet},
  \bibinfo{person}{Felipe~Petroski Such}, \bibinfo{person}{Dave Cummings},
  \bibinfo{person}{Matthias Plappert}, \bibinfo{person}{Fotios Chantzis},
  \bibinfo{person}{Elizabeth Barnes}, \bibinfo{person}{Ariel Herbert-Voss},
  \bibinfo{person}{William~Hebgen Guss}, \bibinfo{person}{Alex Nichol},
  \bibinfo{person}{Alex Paino}, \bibinfo{person}{Nikolas Tezak},
  \bibinfo{person}{Jie Tang}, \bibinfo{person}{Igor Babuschkin},
  \bibinfo{person}{Suchir Balaji}, \bibinfo{person}{Shantanu Jain},
  \bibinfo{person}{William Saunders}, \bibinfo{person}{Christopher Hesse},
  \bibinfo{person}{Andrew~N. Carr}, \bibinfo{person}{Jan Leike},
  \bibinfo{person}{Josh Achiam}, \bibinfo{person}{Vedant Misra},
  \bibinfo{person}{Evan Morikawa}, \bibinfo{person}{Alec Radford},
  \bibinfo{person}{Matthew Knight}, \bibinfo{person}{Miles Brundage},
  \bibinfo{person}{Mira Murati}, \bibinfo{person}{Katie Mayer},
  \bibinfo{person}{Peter Welinder}, \bibinfo{person}{Bob McGrew},
  \bibinfo{person}{Dario Amodei}, \bibinfo{person}{Sam McCandlish},
  \bibinfo{person}{Ilya Sutskever}, {and} \bibinfo{person}{Wojciech Zaremba}.}
  \bibinfo{year}{2021}\natexlab{}.
\newblock \showarticletitle{Evaluating Large Language Models Trained on Code}.
\newblock  (\bibinfo{year}{2021}).
\newblock
\showeprint[arxiv]{2107.03374}~[cs.LG]


\bibitem[Chen et~al\mbox{.}(2023)]%
        {chen2023gotta}
\bibfield{author}{\bibinfo{person}{Xiusi Chen}, \bibinfo{person}{Yu Zhang},
  \bibinfo{person}{Jinliang Deng}, \bibinfo{person}{Jyun-Yu Jiang}, {and}
  \bibinfo{person}{Wei Wang}.} \bibinfo{year}{2023}\natexlab{}.
\newblock \showarticletitle{Gotta: generative few-shot question answering by
  prompt-based cloze data augmentation}. In
  \bibinfo{booktitle}{\emph{Proceedings of the 2023 SIAM International
  Conference on Data Mining (SDM)}}. SIAM, \bibinfo{pages}{909--917}.
\newblock


\bibitem[Cheng et~al\mbox{.}(2023)]%
        {cheng2023prompt}
\bibfield{author}{\bibinfo{person}{Yu Cheng}, \bibinfo{person}{Jieshan Chen},
  \bibinfo{person}{Qing Huang}, \bibinfo{person}{Zhenchang Xing},
  \bibinfo{person}{Xiwei Xu}, {and} \bibinfo{person}{Qinghua Lu}.}
  \bibinfo{year}{2023}\natexlab{}.
\newblock \showarticletitle{Prompt Sapper: A LLM-Empowered Production Tool for
  Building AI Chains}.
\newblock \bibinfo{journal}{\emph{arXiv preprint arXiv:2306.12028}}
  (\bibinfo{year}{2023}).
\newblock


\bibitem[Choi and Lee(2023)]%
        {choi2023codeprompt}
\bibfield{author}{\bibinfo{person}{YunSeok Choi} {and}
  \bibinfo{person}{Jee-Hyong Lee}.} \bibinfo{year}{2023}\natexlab{}.
\newblock \showarticletitle{CodePrompt: Task-Agnostic Prefix Tuning for Program
  and Language Generation}. In \bibinfo{booktitle}{\emph{Findings of the
  Association for Computational Linguistics: ACL 2023}}.
  \bibinfo{pages}{5282--5297}.
\newblock


\bibitem[DePalma et~al\mbox{.}(2024)]%
        {depalma2024exploring}
\bibfield{author}{\bibinfo{person}{Kayla DePalma}, \bibinfo{person}{Izabel
  Miminoshvili}, \bibinfo{person}{Chiara Henselder}, \bibinfo{person}{Kate
  Moss}, {and} \bibinfo{person}{Eman~Abdullah AlOmar}.}
  \bibinfo{year}{2024}\natexlab{}.
\newblock \showarticletitle{Exploring ChatGPT’s code refactoring
  capabilities: An empirical study}.
\newblock \bibinfo{journal}{\emph{Expert Systems with Applications}}
  \bibinfo{volume}{249} (\bibinfo{year}{2024}), \bibinfo{pages}{123602}.
\newblock


\bibitem[Ding et~al\mbox{.}({[n.\,d.]})]%
        {dingcode}
\bibfield{author}{\bibinfo{person}{Xi Ding}, \bibinfo{person}{Rui Peng},
  \bibinfo{person}{Xiangping Chen}, \bibinfo{person}{Yuan Huang},
  \bibinfo{person}{Jing Bian}, {and} \bibinfo{person}{Zibin Zheng}.}
  \bibinfo{year}{[n.\,d.]}\natexlab{}.
\newblock \showarticletitle{Do Code Summarization Models Process Too Much
  Information? Function Signature May Be All What Is Needed}.
\newblock \bibinfo{journal}{\emph{ACM Transactions on Software Engineering and
  Methodology}} (\bibinfo{year}{[n.\,d.]}).
\newblock


\bibitem[Feng and Chen(2023)]%
        {feng2023prompting}
\bibfield{author}{\bibinfo{person}{Sidong Feng} {and} \bibinfo{person}{Chunyang
  Chen}.} \bibinfo{year}{2023}\natexlab{}.
\newblock \showarticletitle{Prompting Is All Your Need: Automated Android Bug
  Replay with Large Language Models}.
\newblock \bibinfo{journal}{\emph{arXiv preprint arXiv:2306.01987}}
  (\bibinfo{year}{2023}).
\newblock


\bibitem[Guo et~al\mbox{.}(2024)]%
        {guo2024exploring}
\bibfield{author}{\bibinfo{person}{Qi Guo}, \bibinfo{person}{Junming Cao},
  \bibinfo{person}{Xiaofei Xie}, \bibinfo{person}{Shangqing Liu},
  \bibinfo{person}{Xiaohong Li}, \bibinfo{person}{Bihuan Chen}, {and}
  \bibinfo{person}{Xin Peng}.} \bibinfo{year}{2024}\natexlab{}.
\newblock \showarticletitle{Exploring the potential of chatgpt in automated
  code refinement: An empirical study}. In
  \bibinfo{booktitle}{\emph{Proceedings of the 46th IEEE/ACM International
  Conference on Software Engineering}}. \bibinfo{pages}{1--13}.
\newblock


\bibitem[Hansson and Ellr{\'e}us(2023)]%
        {hansson2023code}
\bibfield{author}{\bibinfo{person}{Emilia Hansson} {and}
  \bibinfo{person}{Oliwer Ellr{\'e}us}.} \bibinfo{year}{2023}\natexlab{}.
\newblock \bibinfo{title}{Code Correctness and Quality in the Era of AI Code
  Generation: Examining ChatGPT and GitHub Copilot}.
\newblock
\newblock


\bibitem[Husain et~al\mbox{.}(2019)]%
        {husain2019codesearchnet}
\bibfield{author}{\bibinfo{person}{Hamel Husain}, \bibinfo{person}{Ho-Hsiang
  Wu}, \bibinfo{person}{Tiferet Gazit}, \bibinfo{person}{Miltiadis Allamanis},
  {and} \bibinfo{person}{Marc Brockschmidt}.} \bibinfo{year}{2019}\natexlab{}.
\newblock \showarticletitle{Codesearchnet challenge: Evaluating the state of
  semantic code search}.
\newblock \bibinfo{journal}{\emph{arXiv preprint arXiv:1909.09436}}
  (\bibinfo{year}{2019}).
\newblock


\bibitem[Jin et~al\mbox{.}(2024)]%
        {jin2024can}
\bibfield{author}{\bibinfo{person}{Kailun Jin}, \bibinfo{person}{Chung-Yu
  Wang}, \bibinfo{person}{Hung~Viet Pham}, {and} \bibinfo{person}{Hadi
  Hemmati}.} \bibinfo{year}{2024}\natexlab{}.
\newblock \showarticletitle{Can ChatGPT Support Developers? An Empirical
  Evaluation of Large Language Models for Code Generation}.
\newblock \bibinfo{journal}{\emph{arXiv preprint arXiv:2402.11702}}
  (\bibinfo{year}{2024}).
\newblock


\bibitem[Kojima et~al\mbox{.}(2022)]%
        {kojima2022large}
\bibfield{author}{\bibinfo{person}{Takeshi Kojima},
  \bibinfo{person}{Shixiang~Shane Gu}, \bibinfo{person}{Machel Reid},
  \bibinfo{person}{Yutaka Matsuo}, {and} \bibinfo{person}{Yusuke Iwasawa}.}
  \bibinfo{year}{2022}\natexlab{}.
\newblock \showarticletitle{Large language models are zero-shot reasoners}.
\newblock \bibinfo{journal}{\emph{Advances in neural information processing
  systems}}  \bibinfo{volume}{35} (\bibinfo{year}{2022}),
  \bibinfo{pages}{22199--22213}.
\newblock


\bibitem[Lester et~al\mbox{.}(2021)]%
        {power-of-scale}
\bibfield{author}{\bibinfo{person}{Brian Lester}, \bibinfo{person}{Rami
  Al{-}Rfou}, {and} \bibinfo{person}{Noah Constant}.}
  \bibinfo{year}{2021}\natexlab{}.
\newblock \showarticletitle{The Power of Scale for Parameter-Efficient Prompt
  Tuning}. In \bibinfo{booktitle}{\emph{Proceedings of the 2021 Conference on
  Empirical Methods in Natural Language Processing, {EMNLP} 2021, Virtual Event
  / Punta Cana, Dominican Republic, 7-11 November, 2021}},
  \bibfield{editor}{\bibinfo{person}{Marie{-}Francine Moens},
  \bibinfo{person}{Xuanjing Huang}, \bibinfo{person}{Lucia Specia}, {and}
  \bibinfo{person}{Scott~Wen{-}tau Yih}} (Eds.).
  \bibinfo{publisher}{Association for Computational Linguistics},
  \bibinfo{pages}{3045--3059}.
\newblock
\urldef\tempurl%
\url{https://doi.org/10.18653/v1/2021.emnlp-main.243}
\showDOI{\tempurl}


\bibitem[Li et~al\mbox{.}(2023b)]%
        {li2023skcoder}
\bibfield{author}{\bibinfo{person}{Jia Li}, \bibinfo{person}{Yongmin Li},
  \bibinfo{person}{Ge Li}, \bibinfo{person}{Zhi Jin}, \bibinfo{person}{Yiyang
  Hao}, {and} \bibinfo{person}{Xing Hu}.} \bibinfo{year}{2023}\natexlab{b}.
\newblock \showarticletitle{Skcoder: A sketch-based approach for automatic code
  generation}. In \bibinfo{booktitle}{\emph{2023 IEEE/ACM 45th International
  Conference on Software Engineering (ICSE)}}. IEEE,
  \bibinfo{pages}{2124--2135}.
\newblock


\bibitem[Li et~al\mbox{.}(2023c)]%
        {abs-2303-17780}
\bibfield{author}{\bibinfo{person}{Jia Li}, \bibinfo{person}{Yunfei Zhao},
  \bibinfo{person}{Yongmin Li}, \bibinfo{person}{Ge Li}, {and}
  \bibinfo{person}{Zhi Jin}.} \bibinfo{year}{2023}\natexlab{c}.
\newblock \showarticletitle{Towards Enhancing In-Context Learning for Code
  Generation}.
\newblock \bibinfo{journal}{\emph{CoRR}}  \bibinfo{volume}{abs/2303.17780}
  (\bibinfo{year}{2023}).
\newblock


\bibitem[Li et~al\mbox{.}(2023a)]%
        {li2023starcoder}
\bibfield{author}{\bibinfo{person}{Raymond Li}, \bibinfo{person}{Loubna~Ben
  Allal}, \bibinfo{person}{Yangtian Zi}, \bibinfo{person}{Niklas Muennighoff},
  \bibinfo{person}{Denis Kocetkov}, \bibinfo{person}{Chenghao Mou},
  \bibinfo{person}{Marc Marone}, \bibinfo{person}{Christopher Akiki},
  \bibinfo{person}{Jia Li}, \bibinfo{person}{Jenny Chim}, {et~al\mbox{.}}}
  \bibinfo{year}{2023}\natexlab{a}.
\newblock \showarticletitle{StarCoder: may the source be with you!}
\newblock \bibinfo{journal}{\emph{arXiv preprint arXiv:2305.06161}}
  (\bibinfo{year}{2023}).
\newblock


\bibitem[Li and Liang(2021)]%
        {prefix-finetune}
\bibfield{author}{\bibinfo{person}{Xiang~Lisa Li} {and} \bibinfo{person}{Percy
  Liang}.} \bibinfo{year}{2021}\natexlab{}.
\newblock \showarticletitle{Prefix-Tuning: Optimizing Continuous Prompts for
  Generation}. In \bibinfo{booktitle}{\emph{Proceedings of the 59th Annual
  Meeting of the Association for Computational Linguistics and the 11th
  International Joint Conference on Natural Language Processing, {ACL/IJCNLP}
  2021, (Volume 1: Long Papers), Virtual Event, August 1-6, 2021}},
  \bibfield{editor}{\bibinfo{person}{Chengqing Zong}, \bibinfo{person}{Fei
  Xia}, \bibinfo{person}{Wenjie Li}, {and} \bibinfo{person}{Roberto Navigli}}
  (Eds.). \bibinfo{publisher}{Association for Computational Linguistics},
  \bibinfo{pages}{4582--4597}.
\newblock
\urldef\tempurl%
\url{https://doi.org/10.18653/v1/2021.acl-long.353}
\showDOI{\tempurl}


\bibitem[Liu et~al\mbox{.}(2023a)]%
        {liu2023improving}
\bibfield{author}{\bibinfo{person}{Chao Liu}, \bibinfo{person}{Xuanlin Bao},
  \bibinfo{person}{Hongyu Zhang}, \bibinfo{person}{Neng Zhang},
  \bibinfo{person}{Haibo Hu}, \bibinfo{person}{Xiaohong Zhang}, {and}
  \bibinfo{person}{Meng Yan}.} \bibinfo{year}{2023}\natexlab{a}.
\newblock \bibinfo{title}{Improving ChatGPT Prompt for Code Generation}.
\newblock
\newblock
\showeprint[arxiv]{2305.08360}~[cs.SE]


\bibitem[Liu et~al\mbox{.}(2023b)]%
        {liu2023your}
\bibfield{author}{\bibinfo{person}{Jiawei Liu}, \bibinfo{person}{Chunqiu~Steven
  Xia}, \bibinfo{person}{Yuyao Wang}, {and} \bibinfo{person}{LINGMING ZHANG}.}
  \bibinfo{year}{2023}\natexlab{b}.
\newblock \showarticletitle{Is Your Code Generated by ChatGPT Really Correct?
  Rigorous Evaluation of Large Language Models for Code Generation}. In
  \bibinfo{booktitle}{\emph{Thirty-seventh Conference on Neural Information
  Processing Systems}}.
\newblock


\bibitem[Liu et~al\mbox{.}(2024b)]%
        {liu2024your}
\bibfield{author}{\bibinfo{person}{Jiawei Liu}, \bibinfo{person}{Chunqiu~Steven
  Xia}, \bibinfo{person}{Yuyao Wang}, {and} \bibinfo{person}{Lingming Zhang}.}
  \bibinfo{year}{2024}\natexlab{b}.
\newblock \showarticletitle{Is your code generated by chatgpt really correct?
  rigorous evaluation of large language models for code generation}.
\newblock \bibinfo{journal}{\emph{Advances in Neural Information Processing
  Systems}}  \bibinfo{volume}{36} (\bibinfo{year}{2024}).
\newblock


\bibitem[Liu et~al\mbox{.}(2023c)]%
        {2023liucodegen4libs}
\bibfield{author}{\bibinfo{person}{Mingwei Liu}, \bibinfo{person}{Tianyong
  Yang}, \bibinfo{person}{Yiling Lou}, \bibinfo{person}{Xueying Du},
  \bibinfo{person}{Ying Wang}, {and} \bibinfo{person}{Xin Peng}.}
  \bibinfo{year}{2023}\natexlab{c}.
\newblock \showarticletitle{CodeGen4Libs: A Two-Stage Approach for
  Library-Oriented Code Generation}. In \bibinfo{booktitle}{\emph{Proceedings
  of the 38th International Conference on Automated Software Engineering
  (ASE)}}.
\newblock


\bibitem[Liu et~al\mbox{.}(2024a)]%
        {liu2024no}
\bibfield{author}{\bibinfo{person}{Zhijie Liu}, \bibinfo{person}{Yutian Tang},
  \bibinfo{person}{Xiapu Luo}, \bibinfo{person}{Yuming Zhou}, {and}
  \bibinfo{person}{Liang~Feng Zhang}.} \bibinfo{year}{2024}\natexlab{a}.
\newblock \showarticletitle{No need to lift a finger anymore? Assessing the
  quality of code generation by ChatGPT}.
\newblock \bibinfo{journal}{\emph{IEEE Transactions on Software Engineering}}
  (\bibinfo{year}{2024}).
\newblock


\bibitem[Mastropaolo et~al\mbox{.}(2023)]%
        {mastropaolo2023robustness}
\bibfield{author}{\bibinfo{person}{Antonio Mastropaolo}, \bibinfo{person}{Luca
  Pascarella}, \bibinfo{person}{Emanuela Guglielmi}, \bibinfo{person}{Matteo
  Ciniselli}, \bibinfo{person}{Simone Scalabrino}, \bibinfo{person}{Rocco
  Oliveto}, {and} \bibinfo{person}{Gabriele Bavota}.}
  \bibinfo{year}{2023}\natexlab{}.
\newblock \showarticletitle{On the robustness of code generation techniques: An
  empirical study on github copilot}. In \bibinfo{booktitle}{\emph{2023
  IEEE/ACM 45th International Conference on Software Engineering (ICSE)}}.
  IEEE, \bibinfo{pages}{2149--2160}.
\newblock


\bibitem[Nascimento et~al\mbox{.}(2023)]%
        {nascimento2023artificial}
\bibfield{author}{\bibinfo{person}{Nathalia Nascimento}, \bibinfo{person}{Paulo
  Alencar}, {and} \bibinfo{person}{Donald Cowan}.}
  \bibinfo{year}{2023}\natexlab{}.
\newblock \showarticletitle{Artificial intelligence vs. software engineers: An
  empirical study on performance and efficiency using chatgpt}. In
  \bibinfo{booktitle}{\emph{Proceedings of the 33rd Annual International
  Conference on Computer Science and Software Engineering}}.
  \bibinfo{pages}{24--33}.
\newblock


\bibitem[Nguyen et~al\mbox{.}(2020)]%
        {NguyenRRP20}
\bibfield{author}{\bibinfo{person}{Phuong~Thanh Nguyen},
  \bibinfo{person}{Juri~Di Rocco}, \bibinfo{person}{Davide~Di Ruscio}, {and}
  \bibinfo{person}{Massimiliano~Di Penta}.} \bibinfo{year}{2020}\natexlab{}.
\newblock \showarticletitle{CrossRec: Supporting software developers by
  recommending third-party libraries}.
\newblock \bibinfo{journal}{\emph{J. Syst. Softw.}}  \bibinfo{volume}{161}
  (\bibinfo{year}{2020}).
\newblock
\urldef\tempurl%
\url{https://doi.org/10.1016/j.jss.2019.110460}
\showDOI{\tempurl}


\bibitem[Nguyen et~al\mbox{.}(2021)]%
        {apirec}
\bibfield{author}{\bibinfo{person}{Phuong~Thanh Nguyen},
  \bibinfo{person}{Juri~Di Rocco}, \bibinfo{person}{Claudio~Di Sipio},
  \bibinfo{person}{Davide~Di Ruscio}, {and} \bibinfo{person}{Massimiliano~Di
  Penta}.} \bibinfo{year}{2021}\natexlab{}.
\newblock \showarticletitle{Recommending {API} Function Calls and Code Snippets
  to Support Software Development}.
\newblock \bibinfo{journal}{\emph{CoRR}}  \bibinfo{volume}{abs/2102.07508}
  (\bibinfo{year}{2021}).
\newblock
\showeprint[arXiv]{2102.07508}
\urldef\tempurl%
\url{https://arxiv.org/abs/2102.07508}
\showURL{%
\tempurl}


\bibitem[Nijkamp et~al\mbox{.}(2022)]%
        {nijkamp2022codegen}
\bibfield{author}{\bibinfo{person}{Erik Nijkamp}, \bibinfo{person}{Bo Pang},
  \bibinfo{person}{Hiroaki Hayashi}, \bibinfo{person}{Lifu Tu},
  \bibinfo{person}{Huan Wang}, \bibinfo{person}{Yingbo Zhou},
  \bibinfo{person}{Silvio Savarese}, {and} \bibinfo{person}{Caiming Xiong}.}
  \bibinfo{year}{2022}\natexlab{}.
\newblock \showarticletitle{CodeGen: An Open Large Language Model for Code with
  Multi-Turn Program Synthesis}. In \bibinfo{booktitle}{\emph{The Eleventh
  International Conference on Learning Representations}}.
\newblock


\bibitem[Papineni et~al\mbox{.}(2002)]%
        {papineni2002bleu}
\bibfield{author}{\bibinfo{person}{Kishore Papineni}, \bibinfo{person}{Salim
  Roukos}, \bibinfo{person}{Todd Ward}, {and} \bibinfo{person}{Wei-Jing Zhu}.}
  \bibinfo{year}{2002}\natexlab{}.
\newblock \showarticletitle{{BLEU}: a method for automatic evaluation of
  machine translation}. In \bibinfo{booktitle}{\emph{Proceedings of the 40th
  annual meeting of the Association for Computational Linguistics}}.
  \bibinfo{pages}{311--318}.
\newblock


\bibitem[Radford et~al\mbox{.}(2019)]%
        {Radford2019LanguageMA}
\bibfield{author}{\bibinfo{person}{Alec Radford}, \bibinfo{person}{Jeff Wu},
  \bibinfo{person}{Rewon Child}, \bibinfo{person}{David Luan},
  \bibinfo{person}{Dario Amodei}, {and} \bibinfo{person}{Ilya Sutskever}.}
  \bibinfo{year}{2019}\natexlab{}.
\newblock \showarticletitle{Language Models are Unsupervised Multitask
  Learners}.
\newblock
\urldef\tempurl%
\url{https://api.semanticscholar.org/CorpusID:160025533}
\showURL{%
\tempurl}


\bibitem[Reif et~al\mbox{.}(2022)]%
        {ReifIYCCW22}
\bibfield{author}{\bibinfo{person}{Emily Reif}, \bibinfo{person}{Daphne
  Ippolito}, \bibinfo{person}{Ann Yuan}, \bibinfo{person}{Andy Coenen},
  \bibinfo{person}{Chris Callison{-}Burch}, {and} \bibinfo{person}{Jason Wei}.}
  \bibinfo{year}{2022}\natexlab{}.
\newblock \showarticletitle{A Recipe for Arbitrary Text Style Transfer with
  Large Language Models}. In \bibinfo{booktitle}{\emph{{ACL} {(2)}}}.
  \bibinfo{publisher}{Association for Computational Linguistics},
  \bibinfo{pages}{837--848}.
\newblock


\bibitem[Ren et~al\mbox{.}(2020)]%
        {ren2020codebleu}
\bibfield{author}{\bibinfo{person}{Shuo Ren}, \bibinfo{person}{Daya Guo},
  \bibinfo{person}{Shuai Lu}, \bibinfo{person}{Long Zhou},
  \bibinfo{person}{Shujie Liu}, \bibinfo{person}{Duyu Tang},
  \bibinfo{person}{Neel Sundaresan}, \bibinfo{person}{Ming Zhou},
  \bibinfo{person}{Ambrosio Blanco}, {and} \bibinfo{person}{Shuai Ma}.}
  \bibinfo{year}{2020}\natexlab{}.
\newblock \showarticletitle{{CodeBLEU}: a Method for Automatic Evaluation of
  Code Synthesis}.
\newblock \bibinfo{journal}{\emph{arXiv e-prints}} (\bibinfo{year}{2020}),
  \bibinfo{pages}{arXiv--2009}.
\newblock


\bibitem[Roziere et~al\mbox{.}(2023)]%
        {roziere2023codellama}
\bibfield{author}{\bibinfo{person}{Baptiste Roziere}, \bibinfo{person}{Jonas
  Gehring}, \bibinfo{person}{Fabian Gloeckle}, \bibinfo{person}{Sten Sootla},
  \bibinfo{person}{Itai Gat}, \bibinfo{person}{Xiaoqing~Ellen Tan},
  \bibinfo{person}{Yossi Adi}, \bibinfo{person}{Jingyu Liu},
  \bibinfo{person}{Tal Remez}, \bibinfo{person}{J{\'e}r{\'e}my Rapin},
  {et~al\mbox{.}}} \bibinfo{year}{2023}\natexlab{}.
\newblock \showarticletitle{Code llama: Open foundation models for code}.
\newblock \bibinfo{journal}{\emph{arXiv preprint arXiv:2308.12950}}
  (\bibinfo{year}{2023}).
\newblock


\bibitem[Shi et~al\mbox{.}(2022)]%
        {shi2022language}
\bibfield{author}{\bibinfo{person}{Freda Shi}, \bibinfo{person}{Mirac Suzgun},
  \bibinfo{person}{Markus Freitag}, \bibinfo{person}{Xuezhi Wang},
  \bibinfo{person}{Suraj Srivats}, \bibinfo{person}{Soroush Vosoughi},
  \bibinfo{person}{Hyung~Won Chung}, \bibinfo{person}{Yi Tay},
  \bibinfo{person}{Sebastian Ruder}, \bibinfo{person}{Denny Zhou},
  {et~al\mbox{.}}} \bibinfo{year}{2022}\natexlab{}.
\newblock \showarticletitle{Language models are multilingual chain-of-thought
  reasoners}. In \bibinfo{booktitle}{\emph{The Eleventh International
  Conference on Learning Representations}}.
\newblock


\bibitem[Shrivastava et~al\mbox{.}(2023)]%
        {shrivastava2023repository}
\bibfield{author}{\bibinfo{person}{Disha Shrivastava}, \bibinfo{person}{Hugo
  Larochelle}, {and} \bibinfo{person}{Daniel Tarlow}.}
  \bibinfo{year}{2023}\natexlab{}.
\newblock \showarticletitle{Repository-level prompt generation for large
  language models of code}. In \bibinfo{booktitle}{\emph{International
  Conference on Machine Learning}}. PMLR, \bibinfo{pages}{31693--31715}.
\newblock


\bibitem[Siddiq et~al\mbox{.}(2023)]%
        {siddiq2023exploring}
\bibfield{author}{\bibinfo{person}{Mohammed~Latif Siddiq},
  \bibinfo{person}{Joanna Santos}, \bibinfo{person}{Ridwanul~Hasan Tanvir},
  \bibinfo{person}{Noshin Ulfat}, \bibinfo{person}{Fahmid~Al Rifat}, {and}
  \bibinfo{person}{Vinicius~Carvalho Lopes}.} \bibinfo{year}{2023}\natexlab{}.
\newblock \showarticletitle{Exploring the Effectiveness of Large Language
  Models in Generating Unit Tests}.
\newblock \bibinfo{journal}{\emph{arXiv preprint arXiv:2305.00418}}
  (\bibinfo{year}{2023}).
\newblock


\bibitem[Sun et~al\mbox{.}(2019)]%
        {SunXCBW019}
\bibfield{author}{\bibinfo{person}{Jiamou Sun}, \bibinfo{person}{Zhenchang
  Xing}, \bibinfo{person}{Rui Chu}, \bibinfo{person}{Heilai Bai},
  \bibinfo{person}{Jinshui Wang}, {and} \bibinfo{person}{Xin Peng}.}
  \bibinfo{year}{2019}\natexlab{}.
\newblock \showarticletitle{Know-How in Programming Tasks: From Textual
  Tutorials to Task-Oriented Knowledge Graph}. In
  \bibinfo{booktitle}{\emph{{ICSME}}}. \bibinfo{publisher}{{IEEE}},
  \bibinfo{pages}{257--268}.
\newblock


\bibitem[Tunstall et~al\mbox{.}(2022)]%
        {tunstall2022codeparrot}
\bibfield{author}{\bibinfo{person}{Lewis Tunstall}, \bibinfo{person}{Leandro
  Von~Werra}, {and} \bibinfo{person}{Thomas Wolf}.}
  \bibinfo{year}{2022}\natexlab{}.
\newblock \bibinfo{booktitle}{\emph{Natural language processing with
  transformers}}.
\newblock \bibinfo{publisher}{" O'Reilly Media, Inc."}.
\newblock


\bibitem[Vaithilingam et~al\mbox{.}(2022)]%
        {vaithilingam2022expectation}
\bibfield{author}{\bibinfo{person}{Priyan Vaithilingam},
  \bibinfo{person}{Tianyi Zhang}, {and} \bibinfo{person}{Elena~L Glassman}.}
  \bibinfo{year}{2022}\natexlab{}.
\newblock \showarticletitle{Expectation vs. experience: Evaluating the
  usability of code generation tools powered by large language models}. In
  \bibinfo{booktitle}{\emph{Chi conference on human factors in computing
  systems extended abstracts}}. \bibinfo{pages}{1--7}.
\newblock


\bibitem[Vaswani et~al\mbox{.}(2017)]%
        {VaswaniSPUJGKP17}
\bibfield{author}{\bibinfo{person}{Ashish Vaswani}, \bibinfo{person}{Noam
  Shazeer}, \bibinfo{person}{Niki Parmar}, \bibinfo{person}{Jakob Uszkoreit},
  \bibinfo{person}{Llion Jones}, \bibinfo{person}{Aidan~N. Gomez},
  \bibinfo{person}{Lukasz Kaiser}, {and} \bibinfo{person}{Illia Polosukhin}.}
  \bibinfo{year}{2017}\natexlab{}.
\newblock \showarticletitle{Attention is All you Need}. In
  \bibinfo{booktitle}{\emph{{NIPS}}}. \bibinfo{pages}{5998--6008}.
\newblock


\bibitem[Wang and Komatsuzaki(2021)]%
        {gpt-j}
\bibfield{author}{\bibinfo{person}{Ben Wang} {and} \bibinfo{person}{Aran
  Komatsuzaki}.} \bibinfo{year}{2021}\natexlab{}.
\newblock \bibinfo{title}{{GPT-J-6B: A 6 Billion Parameter Autoregressive
  Language Model}}.
\newblock
  \bibinfo{howpublished}{\url{https://github.com/kingoflolz/mesh-transformer-jax}}.
\newblock


\bibitem[Wang et~al\mbox{.}(2022)]%
        {wang2022prompt-tuning-code}
\bibfield{author}{\bibinfo{person}{Chaozheng Wang}, \bibinfo{person}{Yuanhang
  Yang}, \bibinfo{person}{Cuiyun Gao}, \bibinfo{person}{Yun Peng},
  \bibinfo{person}{Hongyu Zhang}, {and} \bibinfo{person}{Michael~R Lyu}.}
  \bibinfo{year}{2022}\natexlab{}.
\newblock \showarticletitle{No more fine-tuning? an experimental evaluation of
  prompt tuning in code intelligence}. In \bibinfo{booktitle}{\emph{Proceedings
  of the 30th ACM Joint European Software Engineering Conference and Symposium
  on the Foundations of Software Engineering}}. \bibinfo{pages}{382--394}.
\newblock


\bibitem[Wei et~al\mbox{.}(2022)]%
        {Chain-of-Thought}
\bibfield{author}{\bibinfo{person}{Jason Wei}, \bibinfo{person}{Xuezhi Wang},
  \bibinfo{person}{Dale Schuurmans}, \bibinfo{person}{Maarten Bosma},
  \bibinfo{person}{Brian Ichter}, \bibinfo{person}{Fei Xia},
  \bibinfo{person}{Ed~H. Chi}, \bibinfo{person}{Quoc~V. Le}, {and}
  \bibinfo{person}{Denny Zhou}.} \bibinfo{year}{2022}\natexlab{}.
\newblock \showarticletitle{Chain-of-Thought Prompting Elicits Reasoning in
  Large Language Models}. In \bibinfo{booktitle}{\emph{NeurIPS}}.
\newblock
\urldef\tempurl%
\url{http://papers.nips.cc/paper\_files/paper/2022/hash/9d5609613524ecf4f15af0f7b31abca4-Abstract-Conference.html}
\showURL{%
\tempurl}


\bibitem[White et~al\mbox{.}(2023)]%
        {white2023chatgpt}
\bibfield{author}{\bibinfo{person}{Jules White}, \bibinfo{person}{Sam Hays},
  \bibinfo{person}{Quchen Fu}, \bibinfo{person}{Jesse Spencer-Smith}, {and}
  \bibinfo{person}{Douglas~C. Schmidt}.} \bibinfo{year}{2023}\natexlab{}.
\newblock \bibinfo{title}{ChatGPT Prompt Patterns for Improving Code Quality,
  Refactoring, Requirements Elicitation, and Software Design}.
\newblock
\newblock
\showeprint[arxiv]{2303.07839}~[cs.SE]


\bibitem[Xu et~al\mbox{.}(2022)]%
        {xu2022polycoder}
\bibfield{author}{\bibinfo{person}{Frank~F Xu}, \bibinfo{person}{Uri Alon},
  \bibinfo{person}{Graham Neubig}, {and} \bibinfo{person}{Vincent~Josua
  Hellendoorn}.} \bibinfo{year}{2022}\natexlab{}.
\newblock \showarticletitle{A systematic evaluation of large language models of
  code}. In \bibinfo{booktitle}{\emph{Proceedings of the 6th ACM SIGPLAN
  International Symposium on Machine Programming}}. \bibinfo{pages}{1--10}.
\newblock


\bibitem[Yusuf et~al\mbox{.}(2022)]%
        {yusuf2022accurate}
\bibfield{author}{\bibinfo{person}{Imam Nur~Bani Yusuf},
  \bibinfo{person}{Lingxiao Jiang}, {and} \bibinfo{person}{David Lo}.}
  \bibinfo{year}{2022}\natexlab{}.
\newblock \showarticletitle{Accurate generation of trigger-action programs with
  domain-adapted sequence-to-sequence learning}. In
  \bibinfo{booktitle}{\emph{Proceedings of the 30th IEEE/ACM International
  Conference on Program Comprehension}}. \bibinfo{pages}{99--110}.
\newblock


\bibitem[Zan et~al\mbox{.}(2022a)]%
        {zan2022language}
\bibfield{author}{\bibinfo{person}{Daoguang Zan}, \bibinfo{person}{Bei Chen},
  \bibinfo{person}{Zeqi Lin}, \bibinfo{person}{Bei Guan}, \bibinfo{person}{Wang
  Yongji}, {and} \bibinfo{person}{Jian-Guang Lou}.}
  \bibinfo{year}{2022}\natexlab{a}.
\newblock \showarticletitle{When Language Model Meets Private Library}. In
  \bibinfo{booktitle}{\emph{Findings of the Association for Computational
  Linguistics: EMNLP 2022}}. \bibinfo{pages}{277--288}.
\newblock


\bibitem[Zan et~al\mbox{.}(2022b)]%
        {zan2022cert}
\bibfield{author}{\bibinfo{person}{Daoguang Zan}, \bibinfo{person}{Bei Chen},
  \bibinfo{person}{Dejian Yang}, \bibinfo{person}{Zeqi Lin},
  \bibinfo{person}{Minsu Kim}, \bibinfo{person}{Bei Guan},
  \bibinfo{person}{Yongji Wang}, \bibinfo{person}{Weizhu Chen}, {and}
  \bibinfo{person}{Jian-Guang Lou}.} \bibinfo{year}{2022}\natexlab{b}.
\newblock \showarticletitle{CERT: Continual Pre-training on Sketches for
  Library-oriented Code Generation}.
\newblock \bibinfo{journal}{\emph{arXiv preprint arXiv:2206.06888}}
  (\bibinfo{year}{2022}).
\newblock


\bibitem[Zhao et~al\mbox{.}(2023)]%
        {zhao2023prompt}
\bibfield{author}{\bibinfo{person}{Shuai Zhao}, \bibinfo{person}{Jinming Wen},
  \bibinfo{person}{Luu~Anh Tuan}, \bibinfo{person}{Junbo Zhao}, {and}
  \bibinfo{person}{Jie Fu}.} \bibinfo{year}{2023}\natexlab{}.
\newblock \showarticletitle{Prompt as Triggers for Backdoor Attack: Examining
  the Vulnerability in Language Models}.
\newblock \bibinfo{journal}{\emph{arXiv preprint arXiv:2305.01219}}
  (\bibinfo{year}{2023}).
\newblock


\bibitem[Zheng et~al\mbox{.}(2023)]%
        {zheng2023codegeex}
\bibfield{author}{\bibinfo{person}{Qinkai Zheng}, \bibinfo{person}{Xiao Xia},
  \bibinfo{person}{Xu Zou}, \bibinfo{person}{Yuxiao Dong},
  \bibinfo{person}{Shan Wang}, \bibinfo{person}{Yufei Xue},
  \bibinfo{person}{Zihan Wang}, \bibinfo{person}{Lei Shen},
  \bibinfo{person}{Andi Wang}, \bibinfo{person}{Yang Li}, {et~al\mbox{.}}}
  \bibinfo{year}{2023}\natexlab{}.
\newblock \showarticletitle{Codegeex: A pre-trained model for code generation
  with multilingual evaluations on humaneval-x}.
\newblock \bibinfo{journal}{\emph{arXiv preprint arXiv:2303.17568}}
  (\bibinfo{year}{2023}).
\newblock


\end{thebibliography}
\bibliographystyle{ACM-Reference-Format}

\end{document}